\documentclass[12pt]{article}
\pdfoutput=1

\usepackage{color}
\usepackage{amsmath}
\usepackage{amssymb}
\usepackage{graphicx}
\usepackage{enumerate}
\usepackage{tensor}
\usepackage{slashed}
\usepackage[multiple]{footmisc}
\usepackage[
      colorlinks=true,
      linkcolor=blue,
      urlcolor=blue,
      filecolor=black,
      citecolor=red,
      ]{hyperref}
\usepackage[normalem]{ulem}
\usepackage{amsfonts}
\usepackage[latin1]{inputenc}
\usepackage{graphicx}
\usepackage{cancel}
\usepackage{mathrsfs}
\usepackage{amssymb}
\usepackage{slashed}
\usepackage{graphicx}
\usepackage{appendix}

\def\tr{\mop{tr}}

\newcommand {\be} {\begin {equation}}
\newcommand {\ee} {\end {equation}}

\newcommand {\bes} {\begin {equation*}}
\newcommand {\ees} {\end {equation*}}

\newcommand{\beq}{\begin{equation}}
\newcommand{\eeq}{\end{equation}}

\def\<{\langle}
\def\>{\rangle}

\def\eg{{\it e.g. }}

\def\({\left(}
\def\){\right)}
\def\[{\left[}
\def\]{\right]}
\def\<{\langle}
\def\>{\rangle}

\def\tr{\mathop{\rm tr}}

\newcommand\half{{\ensuremath{\frac{1}{2}}}}

\newcommand{\KK}{\field{K}}

\newcommand{\TT}{\field{T}}

\newcommand{\bea}{\begin{eqnarray}}
\newcommand{\eea}{\end{eqnarray}}
\newcommand{\bwt}{\begin{widetext}}
\newcommand{\ewt}{\end{widetext}}
\newcommand{\bi}{\begin{itemize}}
\newcommand{\ei}{\end{itemize}}
\newcommand{\ben}{\begin{enumerate}}
\newcommand{\een}{\end{enumerate}}
\newcommand{\bca}{\begin{cases}}
\newcommand{\eca}{\end{cases}}
\newcommand{\bln}{\begin{align}}
\newcommand{\eln}{\end{align}}
\newcommand{\bst}{\begin{split}}
\newcommand{\est}{\end{split}}

\newcommand\ra{{\rightarrow}}

\def\ri{\right}

 \def\be{\begin{equation}}
\def\ee{\end{equation}}
\def\ba#1\ea{\begin{align}#1\end{align}}
\def\bg#1\eg{\begin{gather}#1\end{gather}}
\def\bm#1\em{\begin{multline}#1\end{multline}}
\def\bmd#1\emd{\begin{multlined}#1\end{multlined}}

\def\d{\delta}

\def\k{\kappa}

\def\n{\nu}

\def\la{\label}

\def\({\left(}
\def\){\right)}
\def\[{\left[}
\def\]{\right]}

\usepackage{amsfonts}
\usepackage{amsmath}
\usepackage{url}
\usepackage{hyperref}
\usepackage{bbold}
\usepackage{slashed}
\usepackage{tikz}
\usepackage{graphicx}
\usepackage{epstopdf}
\usepackage{subfigure}
\usepackage{amssymb}
\usepackage{color}
\usepackage{mathrsfs}
\usepackage{fancybox}
\usepackage{lipsum}
\usepackage{eurosym}
\usepackage{simplewick}

\numberwithin{equation}{section}

\usepackage{environ}
\NewEnviron{myequation}{%
\begin{eqnarray}
\scalebox{1.1}{$\BODY$}
\end{eqnarray}
}
\def\d{{\partial}}
\def\n{{\bf \widehat n}}
\def\k{{\bf k}}

\begin{document}
\begin{titlepage}

\setcounter{page}{1} \baselineskip=15.5pt \thispagestyle{empty}

\vfil
\begin{center}

\def\thefootnote{\fnsymbol{footnote}}
\begin{changemargin}{0.05cm}{0.05cm} 
${}$\\[1cm]

\begin{center}
{\large \bf Towards a 2d QFT Analog of the SYK Model}
\end{center} 
\end{changemargin}

~\\[1cm]
{ Gustavo J. Turiaci${}^{\rm a}$ and Herman Verlinde${}^{\rm a,b}$}
\\[0.3cm]

{\normalsize { \sl ${}^{\rm a}$Physics Department and ${}^{\rm b}$Princeton Center for Theoretical Science 
\\[3mm]
Princeton University, Princeton, NJ 08544, USA}}

\end{center}

\vspace{1cm}

 \vspace{0.2cm}
\begin{changemargin}{01cm}{1cm} 
{\small  \noindent 
\begin{center} 
\textbf{Abstract}
\end{center} }
We propose a 2D QFT generalization of the Sachdev-Ye-Kitaev model, which we argue preserves most of its features. The UV limit of the model is described by $N$ copies of a topological Ising CFT. The full interacting model exhibits conformal symmetry in the IR and an emergent pseudo-Goldstone mode that arises from broken reparametrization symmetry. We find that the effective action of the Goldstone mode matches with the 3D AdS gravity action, viewed as a functional of the boundary metric. We compute the spectral density and show that the leading deviation from conformal invariance looks like a $T \bar T$ deformation. We comment on the relation between the IR effective action and Liouville CFT.

\end{changemargin}
 \vspace{0.3cm}
\vfil
\begin{flushleft}
\today
\end{flushleft}

\end{titlepage}

\newpage
\tableofcontents
\newpage

\addtolength{\abovedisplayskip}{.5mm}
\addtolength{\belowdisplayskip}{.5mm}

\def\plus{\raisebox{.5pt}{\tiny$+$\smpc}}

\addtolength{\parskip}{.6mm}
\def\spc{\hspace{1pt}}

\def\nspc{{\hspace{-1pt}}}
\def\ff{\rm\smpc f\smpc} 
\def\fff{\mbox{Y}}
\def\ww{{\rm w}}
\def\smpc{{\hspace{.5pt}}}

\def\zz{{\spc \rm z}}
\def\xx{{\rm x\smpc}}
\def\xxi{\mbox{\footnotesize \spc $\xi$}}
\def\jj{{\rm j}}
 \addtolength{\baselineskip}{-.1mm}

\renewcommand{\Large}{\large}

\def\calO{{b}}
\def\be{\begin{equation}}
\def\ee{\end{equation}}




\def\mathbi#1{\textbf{\em #1}} 
\def\som{{ \textit{\textbf s}}} 
\def\tom{{ \textit{\textbf t}}} 
\def\nom{n} 
\def\mom{m} 
\def\la{\langle}
\def\bea{\begin{eqnarray}}
\def\eea{\end{eqnarray}}
\def\is{\! & \! = \! & \!}
\def\ra{\rangle}
\def\half{{\textstyle{\frac 12}}}
\def\cL{{\cal L}}
\def\halfi{{\textstyle{\frac i 2}}}
\def\ba{\bea}
\def\ea{\eea}
\def\lb{\langle}
\def\rb{\rangle}
\newcommand{\rep}[1]{\mathbf{#1}}

\def\uU{\bf U}
\def\be{\bea}
\def\ee{\eea}
\def\delbar{\overline{\partial}}
\def\ra{\bigr\rangle}
\def\la{\bigl\langle}
\def\ccdot{\!\spc\cdot\!\spc}
\def\nspc{\!\spc\smpc}
\def\tr{{\rm tr}}
\def\ra{\bigr\rangle}
\def\la{\bigl\langle}
\def\li{\bigl|\spc}
\def\ri{\bigr |\spc}

\def\hf{\textstyle \frac 1 2}

\def\bfcdot{\raisebox{-1.5pt}{\bf \LARGE $\spc \cdot\spc $}}
\def\spc{\hspace{1pt}}
\def\is{\! & \! = \! &\!}
\def\d{{\partial}}
\def\n{{\bf \widehat n}}
\def\k{{\bf k}}

\def\pp{{\mbox{\tiny$+$}}}
\def\mm{{\mbox{\tiny$-$}}}

\setcounter{tocdepth}{2}
\addtolength{\baselineskip}{0mm}
\addtolength{\parskip}{.0mm}
\addtolength{\abovedisplayskip}{.5mm}
\addtolength{\belowdisplayskip}{.5mm}

\def\fff{e}

\section{Introduction}\label{sec:intro}
\vspace{-2mm}

The Sachdev-Ye-Kitaev (SYK) model is a soluble quantum many body system that exhibits maximal chaotic behavior \cite{KitaevTalks, Sachdev:2015efa, Polchinski:2016xgd, Maldacena:2016hyu}.\footnote{Similar models have been originally introduced in \cite{Sachdev:1992fk,ParcolletGeorgesnF,Sachdev:2010um,ParcolletGeorges} to model strongly interacting matter systems without quasiparticle excitations that realize non-Fermi liquid states.} It describes the quantum mechanics of $N$ Majorana fermions $\psi_i$ with anti-commutation relations $\{ \psi_i, \psi_j \} = \delta_{ij}$ interacting via a homogeneous non-linear potential with random couplings. The model is specified by the 1D action
\bea
\label{SYK}
 S_{\rm SYK} \is \int \! dt\,\Bigl(  \spc \sum_i \mbox{\large  $\frac{\rm i}{2}$} \psi_i \partial_t \psi_i\, - \, {\rm i}^{\frac q 2}\! \sum_{i_1, ..,i_q} \spc J_{i_1 \dots i_q} \spc \psi_{i_1} \ldots \psi_{i_q}\Bigr)
\ea
Here $J_{i_1...j_q}$ denotes a set of gaussian random couplings. We can split  \eqref{SYK} as $S = S_{\rm UV} + S_{\rm IR}$.
Note that both terms exhibit reparametrization invariance, but that $\psi$ transforms as a scalar in the UV, but has scale dimension $\Delta = 1/q$ in the IR. 
The SYK model exhibits approximate conformal symmetry in the IR, and has been proposed to give a holographic description of a 2D black hole space-time. The link with the gravity dual finds support in the fact that both sides give rise to an effective 1D Goldstone mode whose action is described by the Schwarzian derivative \cite{KitaevTalks,Maldacena:2016hyu, Jevicki:2016bwu, Jevicki:2016ito,Jensen:2016pah,Maldacena:2016upp,Engelsoy:2016xyb, Cvetic:2016eiv}. 

In this note we propose a 2D QFT generalization\footnote{Proposals for 2D generalizations of SYK with a discretized spatial dimension are given in \cite{Gu:2016oyy,Gross:2016kjj,Berkooz:2016cvq}.} of the SYK model \eqref{SYK}, which we argue preserves most of the desired features. In particular, via the same reasoning that applies to 1D case, we will argue that the 2D model appears to exhibit conformal symmetry in the IR and gives rise to an emergent Goldstone mode associated with broken 2D reparametrization invariance. We find that the same effective action of the Goldstone mode can also be derived from the 3D AdS gravity action, viewed as a functional of the boundary metric. These results indicate that our 2D model flows in the IR to a holographic 2D CFT, and may thus provide new insight into the dynamical mechanism that underlies AdS${}_3$/CFT${}_2$ duality.

Recently, Witten found an adaptation of a class of so-called tensor models that give rise to the same large $N$ diagrammatical rules as the SYK model \cite{Witten:2016iux}\cite{igorgrisha}. 
It would be worthwhile to investigate whether our proposal can be adapted to this case.

This paper is organized as follows. In section \ref{sec:Model}, we specify our 2D model. We give both a Lagrangian and  Hamiltonian formulation. We give special attention to the UV limit, which is described by a topological Ising CFT. In section \ref{sec:SD} we study the Schwinger-Dyson equations that capture the large $N$ dynamics of the model. We describe the solution of the SD equations in the conformal IR regime, and study the four point function. We find that the chiral spectrum of our 2D model coincides with that of the 1D SYK model. In section \ref{sec:EffectiveAction}, we analyze the dynamics of the pseudo-Goldstone mode, and show that its action is given by the product of two Schwarzian derivatives. We then exhibit the connection with 2+1-D gravity. In section \ref{sec:Conclusion} we list some open questions. In the Appendix \ref{sec:App} we summarize the properties of topological RCFTs.

\def\YY{\mbox{$\phi$}}
\def\ppm{{\mbox{\tiny $\pm$}}}
\def\XX{\mbox{\small $X$}}
\def\UU{\mbox{\small $U$}}
\def\VV{\mbox{\small $V$}}
\section{The 2D model}\label{sec:Model}

\vspace{-2mm}

In this section, we will give two characterizations of our 2D model. First we introduce the model via its Lagrangian, and then we present a Hamiltonian formulation. 
We give some special attention to the UV limit of our model.
 
\def\beps{\mbox{\large $\!\spc\epsilon$}}

\subsection{Lagrangian formulation}\label{sec:ModelLag}
 
\vspace{-2mm}

A na\"ive attempt to generalize the SYK model to 2D is to promote the $\psi$ variables to 2D Majorana fermions with a standard kinetic term $\frac{\rm i}{2}\psi\slashed{\partial}\psi$. This choice assigns canonical scale dimension $[\psi]=1/2$. The interaction term then has dimension $q/2$, which is at best marginal. In the 1D action \eqref{SYK}, on the other hand, the UV term assigns $\psi$ scale dimension $[\psi]= 0$, so the interaction term is relevant and the model is strongly coupled in the IR. 

\def\la{\bigl\langle}
\def\ra{\big\rangle}

To write the 2D generalization of \eqref{SYK} we introduce fermionic variables $\psi^i_+$ and $\psi_-^i$ with $i=1, ..., N$.
One can think of $\psi_+$ and $\psi_-$ as the two chiral components of a 2D Majorana fermion. However, to preserve the essential features of the SYK dynamics, we replace the usual fermion kinetic term by the UV term in the following 2D action
\bea
&& \qquad \quad S \; = \; S_{\rm UV} \, + \, S_{\rm IR} \nonumber\\[4mm]
&& \hspace{-5mm} S_{\rm UV}\,=\,  \sum_i \int\!\!\spc d^2 {\rm x}\, \epsilon^{\mu\nu}  \spc\psi^i_+ \partial_\mu \psi^i_+\, \psi^i_- \partial_\nu \psi^i_-
 \label{action} \\[.5mm]
S_{\rm \spc IR}\is\!\!
\sum_{i_1,...,j_q}\,
\int \!\!\spc d^2{\rm x} \; J_{i_1\spc.\spc.\spc.\spc 
j_q}\; \psi^{i_1}_-\spc 
\spc \ldots \, \psi^{i_q}_-\;  \psi^{j_1}_+\spc 
\spc \ldots \, \psi^{j_q}_+\, \nonumber
\ea
where $J_{i_1...j_q}$ denote a set of gaussian random couplings with 
\bea
\label{gaussjt}
\qquad \bigl\langle (J_{i_1...j_q})^2 \bigr\rangle \is \frac{J^2 \spc (q\!\spc-\!\spc 1)! \spc q!}{N^{2q-1}} ~~~~~\text{ (no sum) }.
\eea
The unconventional kinetic term\footnote{The quartic kinetic term in \eqref{action} can be viewed as a fermionic cousin of the Nambu-Goto action. It is also similar to the  fermionic Wess-Zumino term that appears in the Green-Schwarz superstring action \cite{gs}.}  in $S_{\rm UV}$ is chosen such that $\psi$ has canonical scale dimension $[\psi]_{{}_{\rm UV}} = 0$.  The couplings in $S_{\rm IR}$ thus have dimension $[J]=2$. The interaction term is therefore relevant and dominates the IR dynamics.

The total action defines a proper relativistic QFT, but does not come with a fixed light cone. Both terms in \eqref{action} do not depend on a choice of metric: the UV term is topological, whereas the IR term only requires a choice of integration measure.
$S_{\rm UV}$ is reparametrization invariant if $\psi_\pm$ transform as scalars, while
 $S_{\rm IR}$ has reparametrization symmetry provided the fermions transform as 
$\psi_{a}^{i}(\xx)  \to  \bigl|\det \frac{\partial \tilde{\rm x}^\mu}{\partial {\rm x}^\nu}\bigr|{}^{1/{2q}}\, \psi'_a{\!}^{i}(\tilde{\xx}(\xx))$. The fact that the UV and IR transformation laws are different is a first hint that the model may give rise to an effective Goldstone mode associated with broken reparametrization symmetry. The UV and IR action still share area preserving diffeomorphisms as a common symmetry group. 

Note that the quartic kinetic term involves a diagonal pairing  between the chiral partners $\psi_+^i$ and~$\psi_-^i$, but the IR interaction term does not.
The action \eqref{action} is invariant under local Lorentz transformations $\psi_\pm^i\to \lambda^{\pm1}\psi_\pm^i$. For the UV action, these can act independently on each sector. We will treat the overall local Lorentz invariance as a gauge symmetry. 

The quartic kinetic term is a central new ingredient of our proposal. So it is important to understand its physical role and consequences. We have seen some of its desirable properties. Some apparent draw backs are that it obscures the form of the anti-commutation relations and does not produce a standard  fermion propagator. 
To gain some further insight, let us take a closer look at the theory defined by $S_{\rm UV}$ just by itself.

\subsection{UV limit: Topological Ising CFT}\label{sec:ModelTICFT}
\vspace{-1mm}

The UV theory splits up into $N$ decoupled topological theories with a single pair of chiral Majorana fermions each. Let us focus on one of these UV sectors. 
 A non-linear fermionic action 
similar to $S_{\rm UV}$ with $N=1$ was recently considered in \cite{hansson} in the context of a proposed topological theory of Majorana edge modes of a $p_x+ip_y$ superconductor.\footnote{A similar topological fermionic model has also been considered by D. Haldane (private communication).}

By introducing Hubbard-Stratonovich variables $e^\pm_\mu$  we can rewrite the UV action as
\bea
\label{hsaction}
S \is\frac 1 2   \int \! d^2 {\rm x} \, \epsilon^{\mu\nu} \bigl(\spc e^{a}_\mu \, \psi_a\spc \partial_\nu \psi_a \spc - \spc \spc \epsilon_{ab}\spc e^{a}_\mu \; e^{b}_\nu \; \bigr).
\ea
with $a=\pm$.
This action is manifestly reparametrization and local Lorentz  invariant.
 We can think of  the $e^{a}_\mu$ variable as a Cartan zweibein, that parametrizes a dynamical 2D metric and local Lorentz frame. For fixed $e^a_\mu$, the action \eqref{hsaction} has a conventional fermion kinetic term. Integrating out $e_\mu^a$ gives back the quartic action.

Let us take a brief look at the classical theory. The equations of motion of \eqref{hsaction} imply 
\bea
\epsilon^{\mu\nu} e^+_\mu \spc \psi_+\partial_\nu \psi_+ \! \is  0, \qquad \qquad 
 e^+_\mu \, = \, \psi_- \partial_\mu \psi_-,\nonumber\\[-2mm]
\label{uveom} \\[-2mm]
\epsilon^{\mu\nu} e^-_\mu \psi_-\partial_\nu \psi_-\! \is 0, \qquad \qquad e^-_\mu \, = \, \psi_+ \partial_\mu \psi_+ .\nonumber
\eea
Locally we can introduce two scalar fields $\XX^\pm$ such that
\bea
\label{xpmd}
e^+_\mu\! \is e^{\varphi_+} \partial_\mu \XX^+ , \qquad\quad e^-_\mu \, = \, e^{\varphi_-}  \partial_\mu \XX^- . 
\eea
We can then solve the equation of motion \eqref{uveom} by setting $\psi_-(\XX^-)$ and $\psi_+(\XX^+).$
So  for a moment  it looks like $\psi_-$ and $\psi_+$ behave like a pair of chiral fermions that propagate along two independent light-cone directions specified by $\XX^-$ and $\XX^+$. However, from \eqref{uveom} and \eqref{xpmd} we also deduce that
\bea
\epsilon^{\mu\nu} \partial_\mu \XX^+ \partial_\nu \XX^- \, = \, 0
\eea
which can only be solved if the two light-cone directions in fact coincide. So the UV model \eqref{hsaction} does not have true propagating modes. As we will argue below, it describes  a topological field theory.

Introducing a dynamical 2D metric via $g_{\mu\nu} = \eta_{ab}\spc e^a_\mu e^b_\nu$, assembling $\psi_+$ and $\psi_-$ into a two component fermion, and performing a simple field rescaling $\tilde\psi = g^{1/8} \psi$,
we may further rewrite \eqref{hsaction} as a standard action of a 2D Majorana fermion coupled to 2D gravity
\bea
\label{gravaction}
S \is  \int \! \sqrt{g} \bigl(\spc \mbox{\large $\frac {\rm i} 2$} \tilde\psi \spc \slashed\nabla \tilde\psi  
\spc - \spc \spc 1 \; \bigr).
\ea
This rewriting of $S_{\rm UV}$ is closely analogous to the procedure that recasts the Nambu-Goto action into that of a free boson coupled to 2D gravity.
Minimal models coupled to 2D gravity have been studied extensively, starting with KPZ \cite{KPZ}. Our treatment will need to be somewhat different.  In the end, we want to be able to add the IR action in \eqref{action} as an  interaction term. Since the interaction term is invariant only under area preserving diffeomorphisms, we are not allowed to treat the full diffeomorphism group as a gauge symmetry of the UV theory. So instead of viewing the model as a gravitational theory,  we will treat it as a topological CFT with local gauge invariant observables \cite{robbert,HV,WZW-gauged, BlauThompson}. 

Equations \eqref{hsaction} and \eqref{gravaction} describe a 2D Ising CFT with gauged Virasoro symmetry. The gauging projects out all Virasoro descendent and leaves only three local observables given by the dressed primary operators: the unit operator~$\mathbb{1}$,   
 the spin field $\sigma$ and 
\bea
\beps(\xx) \is \psi_+(\xx)\psi_-(\xx).
\ea
We will call this theory  the topological Ising CFT. It is the simplest example of a topological RCFT.  Some relevant properties of topological RCFTs are  summarized in the Appendix. For the purpose of our main discussion here, it is sufficient to note that:

\smallskip

\noindent
$\bullet$ In Euclidean signature, the correlation functions of local observables are independent~of ${}$~~positions of the operators. They equal an integer, given by the number of independent ${}$~~{\spc}chiral conformal blocks associated with the corresponding CFT correlation function \cite{BlauThompson}.\\
$\bullet$ The three gauge invariant local observables $\beps$, $\sigma$ and $\mathbb{1}$ all have scale dimension zero. The  ${}$~~\spc\,operator algebra forms a commutative, associative ring isomorphic to the Ising fusion~rules
$$
\mathbb{1} \times\mathbb{1} = \mathbb{1}, \quad \ \mathbb{1} \times \sigma = \sigma, \quad \ \mathbb{1} \times \beps = \beps,\quad \ \beps \times \beps = \mathbb{1},  \quad \ \beps \times \sigma = \sigma, \quad \ \sigma \times \sigma = \mathbb{1} + \beps.
$$
$\bullet$ In Minkowski space, TCFT correlation functions acquire non-trivial position dependence ${}$~~\;due to operator ordering. This dependence reflects the monodromy of the chiral conformal ${}$~~\,blocks, or equivalently, the topological braid properties of the chiral CFT operators.

\smallskip

Let us elaborate this last point. Just like in an ordinary RCFT, local observables in a topological RCFT can be factorized into a sum of chiral components. The properties of these chiral components is made most manifest by formulating the TCFT as a gauged WZW model \cite{BlauThompson}. In this formulation, the chiral operators are attached to a Wilson line of a flat gauge field that stretches out to the corresponding (past or future) null infinity. The Wilson lines encode the topological braid properties of the chiral operators of the CFT.

\begin{figure}[t]
\begin{center}
\includegraphics[scale=0.6]{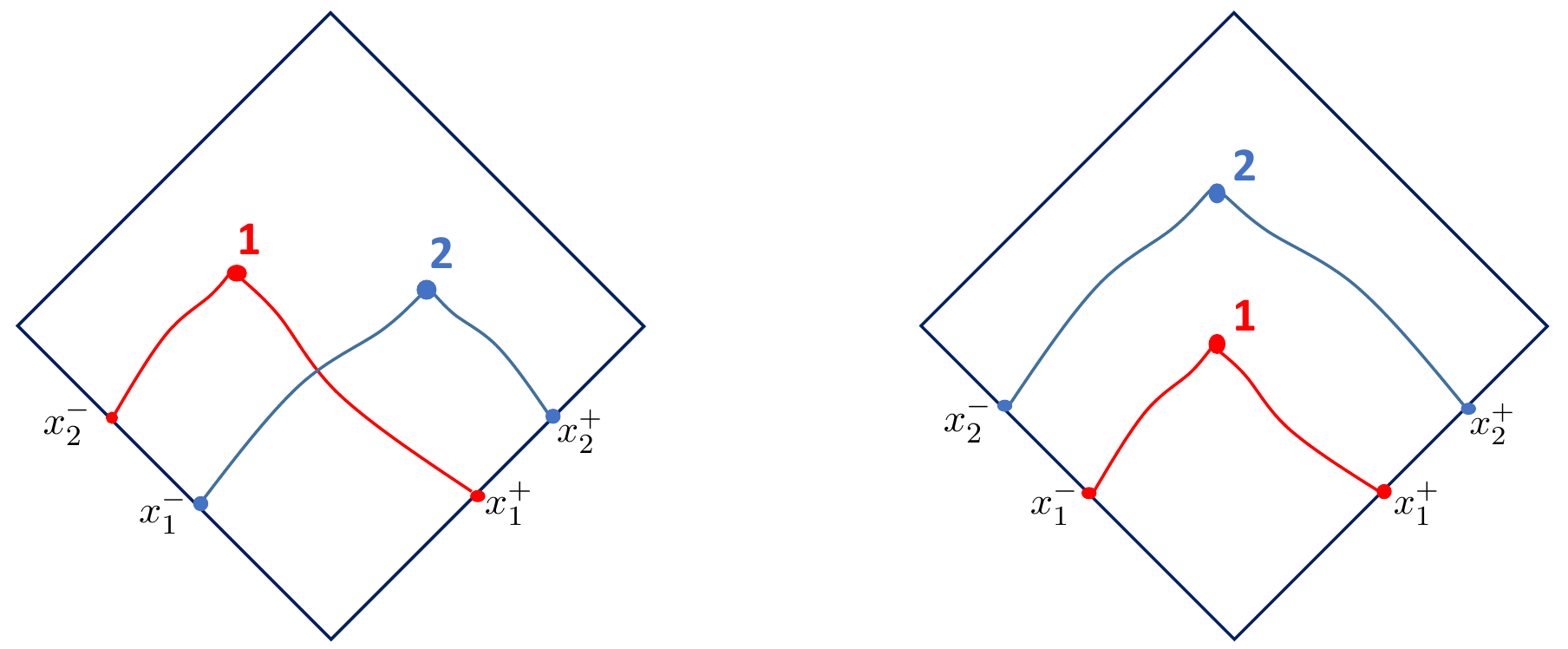}
\caption{\label{wfactor} {\small \it In a topological CFT, local operators are attached to two Wilson lines that connect to past null infinity. Whether two operators are space-like or time-like separated is a topological distinction, encoded via the relative ordering of the asymptotic end-points $x^\pm_1$ and $x^\pm_2$ of the respective Wilson lines. The lines are shown with zigzags to indicate that the bulk has no fixed metric. }}
\vspace{-5mm}
\end{center}
\end{figure}

Hence local operators in a TCFT look as indicated in figure \ref{wfactor}.  The space-time position of a local operator is labeled by the locations $x_i^+$ and $x_i^-$ where the Wilson lines attach to past null infinity. Since null infinity of 2D Minkowski space-time is one-dimensional, time ordering again becomes topological.
The 2D light-cone thus also becomes a topological notion, that divides 2D space-time into four regions. Correspondingly, for each pair of operators we can distinguish four types of relative separations:  past, future, left and right. Thanks to the presence of the Wilson lines, these four are all topologically distinct.

\def\Ssigma{{\mbox{\small $\Sigma$}}}

Specializing to the simplest example: for the 2-point function of two $\beps$ operators in the topological Ising model, the prescription outlined above and in the Appendix reduces to 
\bea
\la\spc \mathbb{T} \spc \, \beps(1) \, \beps(2)\, \ra_{{ \!\spc}_{\rm TCFT}}\! \is \mbox{$\left\{ \hspace{-2mm}  \begin{array}{cc}{\langle \psi_+\!\spc(1)\psi_+\!\spc(2)\rangle\langle \psi_-\!\spc(1)\psi_-\!\spc(2)\rangle \, =\; 1}&\ {\rm F}\\[.7mm] {\ \; \langle \psi_+\!\spc(2)\psi_+(1) \rangle\langle\psi_-\!\spc(1)\psi_-\!\spc(2)\rangle\, = -1} &\ {\rm R} 
\\[.7mm] {\ \; \langle \psi_+\!\spc(1)\psi_+\!\spc(2)\rangle\langle \psi_-\!\spc(2)\psi_-\!\spc(1)\rangle\, = -1} &\ {\rm L} \\[.7mm] {\langle \psi_+\!\spc(2)\psi_+\!\spc(1)\rangle\langle \psi_-\!\spc(2)\psi_-\!\spc(1)\rangle\, =\; 1}&\  {\rm P} \end{array}\right.$}\raisebox{-1cm}{\qquad \qquad\ $$\includegraphics[scale=0.43]{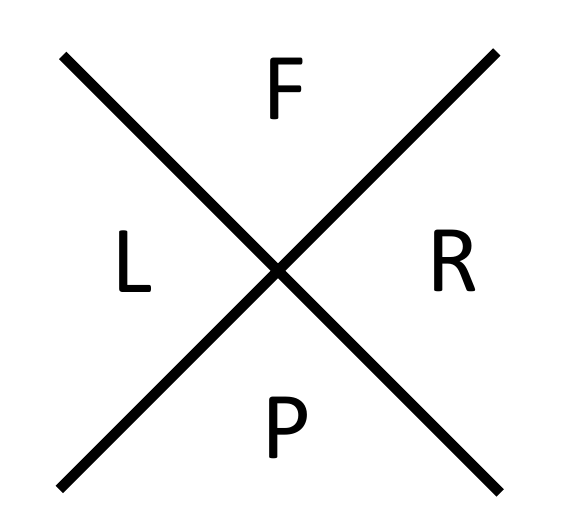}$$}\nonumber
\eea
The four outcomes correspond to four different operator orderings. Here we introduced the double time ordering symbol ${\mathbb T} ={\rm P}_+{\rm P_-}$, where ${\rm P}_\pm$ denotes the time ordering symbol that orders the operators according to increasing light-cone time $\pm x^\pm$. We can abbreviate the above table as
\bea
\label{etwo}
\la \mathbb{T} \, \beps({\rm x}_1) \, \beps({\rm x}_2)\ra_{{ \!\spc}_{\rm TCFT}} \!\! \is {\rm sgn}(x^+_{12}) \, {\rm sgn}(x^-_{12})
\eea
Here the 2D location  ${\rm x} = (x^+,x^-)$ of each operator is defined via the position of the end-points of the Wilson lines, as indicated in figure \ref{wfactor}.
The formula \eqref{etwo} should be compared with the formula $\langle {\rm T}\spc \psi(\tau_1) \psi(\tau_2)\rangle = {\rm sgn}(\tau_{12})$ for the 2-point function of a single free Majorana fermion.\footnote{The vacuum two-point function of free 1D Majorana fermions remains unchanged at finite temperature  \cite{KitaevTalks,Maldacena:2016hyu}.  The same property holds true for the vacuum two-point function in our topological UV theory. This statement would not be true for 2D Majorana fermions with the usual kinetic term. } 
 It forms the basis for the rest of our story.
 
More generally, applying the TCFT rules to the $n$-point function gives that
\bea
\label{npt}
 \bigl\langle\, \mathbb{T} \, \beps(1) \, \beps(2) \spc ...\; \beps(n) \bigr\rangle_{{ \!\spc}_{\rm TCFT}}\!\!  \is \, \left\{ 
 \begin{array}{cc}(-1)^{\#(1,2,...,\spc n)} & \quad \mbox{ $n$ even} \\[-1mm]
0 &\quad \mbox{$n$ odd} \end{array} \right.
\eea
where $\#(1,2,...,\spc n)$ counts the number of times a pair of operators needs to cross each other's light cone in order to rearrange all operators to be space-like separated. Note that, since $\beps$ and $\mathbb{1}$ have a unique OPE channel, at most one single chiral conformal block contributes for each $n$-point function. So the value of the Euclidean $n$ point function is simply equal to 1. 

The expression \eqref{npt} can be rewritten in somewhat more familiar form as follows 
\bea
\label{nptpf}
 \bigl\langle\, \mathbb{T} \, \beps({\rm x}_1) \spc ...\; \beps({\rm x}_n) \bigr\rangle_{{ \!\spc}_{\rm TCFT}}\!\!  \is {\rm Pf}\bigl({\rm sgn}(x^+_{ij})\bigr)\,  {\rm Pf}\bigl({\rm sgn}(x^-_{ij})\bigr) .
\eea
A proof of the equality between \eqref{npt} and \eqref{nptpf} is given in the Appendix.
We see that the non-chiral $n$-point functions factorize into a product of two chiral factors. This factorization property allows us to define the $n$-point functions of the chiral Majorana fermions as
\bea
\label{mapf}
 \bigl\langle\, \mathbb{T} \, \psi_\pm({\rm x}_1) \spc ...\; \psi_\pm({\rm x}_n) \bigr\rangle_{{ \!\spc}_{\rm TCFT}}\!\!  \is {\rm Pf}\bigl({\rm sgn}(x^\pm_{ij})\bigr)\, 
\eea
It is natural to refer to the chiral fields $\psi_+(x^+)$ and $\psi_-(x^-)$ as `topological 2D Majorana-Weyl fermions'. They arise from the topological Ising model after performing a chiral~projection. 

Our 2D model \eqref{action} in fact makes essential use of a chiral projection of this kind. Each term in the interaction Lagrangian in \eqref{action} contains an equal number of left- and right chiral fermions, but the pairing can be off diagonal. In other words, the interaction term is built up from general fermion bi-linears $\psi_+^i \psi_-^j$. To allow for such operators with $i\neq j$, while preserving locality, we need to perform an analog of the GSO projection familiar from superstring theory.
The complete UV theory is defined by taking a tensor product of $N$ topological Ising models, and then performing a chiral projection that allows us to act with general fermion bi-linears $\psi_+^i \psi_-^j$. Similar to the GSO projection, this eliminates the non-chiral spin operators $\sigma_i$ of each individual topological Ising model from the UV spectrum.  The resulting theory then remains local.

\subsection{Hamiltonian formulation}\label{sec:ModelHamiltonian}

\vspace{-1mm}

We would like to verify that the 2D action \eqref{action} defines a unitary QFT. The Hamiltonian formulation is usually most well adapted for this purpose.
So let us write ${\rm x} = (x,t)$ and 
identify the Hilbert space ${\cal H}$ of states defined on a constant time-slice. We should then check that there are no negative norm states and that the Hamiltonian generates a unitary time evolution. The formalism of matrix product states \cite{mps} will turn out to be helpful.

In many interesting quantum many body systems, the wave function $|\Psi\rangle$ depends in a non-trivial way on 
on the spatial ordering of the quasi-particles. A matrix product state (MPS) representation of a quantum state encodes this spatial dependence  by means of an auxiliary quantum system \cite{mps}. 
To define this auxiliary quantum system for our setting, we introduce two collections of $N$ Majorana~fermions with anti-commutation relations
\bea
\{ \psi_\pm^i(x),\psi_\pm^j(y) \} \! \is\! \spc \delta^{ij}, 
\qquad \quad \{\psi^i_+(x),\psi^j_-(y)\}\,=\, 0,
\eea
with $i,j=1,..,N$.
Note that the anti-commutator does not depend on the locations $x$ and $y$. So we can simply set $\psi_\pm^i(x) = \psi_\pm^i$ with $\{ \psi_\pm^i,\psi_\pm^j \} = \spc \delta^{ij}$ acting on a $2 \times 2^{N/2}$ dimensional auxiliary Hilbert space. The role of the position $x$ is to keep track of spatial operator ordering within the matrix product state, in the same way that time $t$ can be used to keep track of time ordering for a free 1D Majorana fermion.

States in the Hilbert space ${\cal H} = {\cal H}_+\otimes {\cal H}_-$ are given by a sum of factorized states  $|\Psi_+^I \rangle \spc | \Psi_-^J\rangle$
where $I$ and $J$ represent a multi-index, e.g. $I = \{{i_1,...,i_p}\}$ labeling the internal quantum numbers of the chiral Majorana particles. Each factor
$|\Psi_\pm^I\rangle$ is represented by a many body wave function in the form of a matrix product state 
\bea
 \Psi^{i_1...i_p}_\pm (x^\pm_{i_1}, ... \spc ,x^\pm_{i_p}) \is
  \la n_\pm \bigr|\spc {\rm P}_{\! \pm} \,\spc  {\psi}_\pm^{i_p}(x^\pm_{i_p})\spc \ldots\, {\psi}_\pm^{i_1}(x^\pm_{i_1}) \spc\bigl|0\ra 
\eea
where $|n_\pm\rangle$ is short-hand for the unique fermion number eigen state that gives a non-zero overlap. Here ${\rm P}_{\! \pm}$ denotes the path ordering symbol that places the operators in spatial order with position $x^\pm_k$ increasing from left to right for $\Psi_-$ and from right to left for $\Psi_+$. Alternatively, we can write the MPS wave function as a 1D path integral 
\bea
 \Psi^{i_1...i_p}_\pm (x^\pm_{i_1}, ... \spc ,x^\pm_{i_p})   \is  \int [d\psi^i_\pm]\, e^{\spc \mbox{\footnotesize $\pm \sum_i \int\! dx^\pm \spc\frac{\rm i} 2 \psi_\ppm^i \partial_\ppm \psi_\pm^i$}}  \; {\psi}_\pm^{i_p}(x^\pm_{i_p})\spc \ldots \spc {\psi}_\pm^{i_1}(x^\pm_{i_1})\nonumber
\eea
Note that this functional integral is reparametrization invariant in $x^\pm$, and that $\Psi^I_\pm$ is a piece-wise constant function of the positions $x^\pm_{i_k}$. In the case that all $\psi^i$'s have the same index, it reduces to the Pfaffian expression \eqref{mapf}.

The Hamiltonian of our model is defined as a linear mapping on the MPS wave functions. It is given by a pure interaction term
$H = \hat{H}_{\rm int}(t) = -\int \! dx \, \hat{\cal L}_{int}(x,t)$ with
\bea
\hat{\cal L}_{\rm int}(x,t) \is \sum_{i_1,..,j_q}\! J_{i_1...j_q}\; \hat\psi_+^{i_1}(x,t) \,  \ldots \, \hat\psi_-^{j_q}(x,t)
\eea
the same interaction term as in \eqref{action}, and where $\hat{\psi}{\spc}^i_\pm(x^\pm)$ with $x^\pm = x\pm t$ now denote  operators that insert $\psi^i_\pm(x^\pm)$ into the corresponding chiral MPS wave function.  Here we reintroduced the time dependence as prescribed by the interaction picture. Note, however, that the free Hamiltonian $H$ identically vanishes.  The $t$ dependence is therefore spurious, except for its effect on operator ordering. The dependence on the two light cone coordinates $x^\pm$ arises due to the intrinsic path-ordering of the matrix product states. 

Integrating the Schr\"odinger equation produces a double lightcone-time ordering prescription
\bea
{\rm T}
\exp\Bigl(-{\rm i}\!\int \! dt \, \hat{H}_{\rm int}(t)\Bigr)
 \is \mathbb{T}
\spc 
\exp\left({\rm i}\!\int \! dx^\pp \!\! \int \! dx^\mm  \hat{\cal L}_{\rm int}(x^\pp,x^\mm)\right)
\eea
where $\mathbb{T} \equiv {\rm P}_{\! \pp}{\rm P}_\mm$ puts all operators in order of increasing light cone time, both along the $x^+$ and  $-x^-$ direction. In this way, through the use of the matrix product state formalism,  we have made contact with the TCFT prescription outlined in the previous subsection.

The last remaining task is to provide an inner product on ${\cal H}$. It seems reasonable to assume that it can be defined such that the states $|\Psi_\pm^I\rangle$ form an orthonormal basis of the respective chiral Hilbert spaces~${\cal H}_\pm$. In principle one should be able to derive this inner product from the path-integral formulation, starting from the action \eqref{action}, or vice versa, derive the path-integral and the action \eqref{action} from the Hamiltonian formalism outlined here. We leave this problem for future study.

\def\xx{{\rm x}}
\section{Schwinger-Dyson equations}\label{sec:SD}

\vspace{-1mm}

Now that we have introduced the 2D model, we would like to analyze its large $N$ dynamics. The factors of $N$ in (\ref{gaussjt}) are chosen so that the model admits a regular large $N$ limit.
We would like to analyze the low point correlation functions of the  2D model, working to leading order in $1/N$.
Throughout, we will assume that the standard SYK analysis applies to our 2D model. In particular, we assume that we can use the replica method to take the disorder average, and that the model does not undergo a spin glass transition.
 
 \subsection{SD equations at large $N$}
 
The simplest non-trivial correlation function with a regular large $N$ limit is
\bea
\label{ffour}
{\cal F}(\xx_1,\xx_2) \is \frac{1}{N^2} \,\sum_{i,j}\, \la \psi^i_+ (x_1)\psi^j_-(x_1) \psi^i_+ (x_2) \psi_-^j(x_2)\ra \nonumber
\ea
At leading order in $1/N$, it factorizes as
\bea
\label{factorize}
{\cal F}(\xx_1,\xx_2) \is\, G_+(\xx_1,\xx_2)\spc G_-(\xx_1,\xx_2) 
\eea
We can identify $G_\pm(\xx_1,\xx_2)$ with the dressed fermionic two point functions
\bea 
\label{gtwo}
G_\pm(\xx_1,\xx_2) \is \frac{1}{N} \, \sum_i\, \la \spc \psi^i_\pm (\xx_1) \psi^i_\pm (\xx_2)\spc \ra,
\eea
with the understanding that each should always appear in the local combination \eqref{factorize}.

To compute the two point functions, we can try to follow the standard SYK procedure and sum all relevant leading order diagrams.  We start by writing the UV action  in the Hubbard-Stratonovich form (\ref{hsaction}) by introducing a total of $N$ Cartan frames $e^\pm_i$, one for each of the $N$ sectors. It is not difficult to see, however, that by restricting ourselves to observables of the type \eqref{ffour} and \eqref{gtwo}, defined as equal weighted sums over all $N$ sectors, that only the collective field
\bea
\label{eav}
e^\pm \is \frac 1 N\; \sum_i \, e^\pm_i
\eea
participates in the large $N$ dynamics. More precisely, if we split each frame variable as $e^\pm_i = e^\pm + \tilde{e}^\pm_i$, the deviation $\tilde{e}^\pm$ will decouple in correlation functions of averaged observables. This property follows from the fact that the interaction term between frame variables and $\psi_\pm^i$ is linear in $e^\pm_i$, and that the fermion propagator lines involve a uniform sum over $i$. So the frame variables~always~couple~via\\[2.5mm]
${}$~~~~~~~~~~~~~~~~~~~~~~~~~~~~~~~\includegraphics[scale=.65]{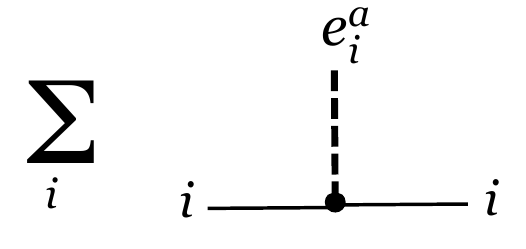}~~\raisebox{2pt}{$\includegraphics[scale=.65]{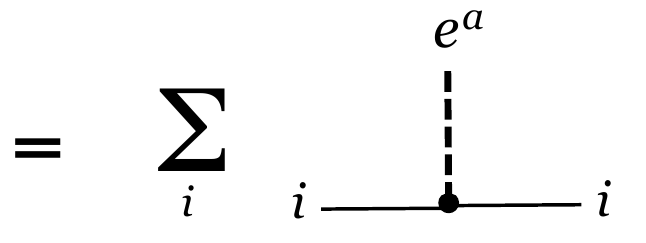}$}\\[2.5mm]
For the computation of large $N$ correlation functions, we can thus replace the frame variables by their large $N$ average \eqref{eav} and use the following effective form of the UV action
\bea
\label{hsactiont}
{S_{\rm UV}}  \is \frac 1 2  \int \! d^2 x \, \epsilon^{\mu\nu} \spc \Bigl(\, \sum_{i,a}  e^a{\!\!}_\mu\, \psi_a^i \partial{}_\nu \psi_a^i - N e^+_\mu e^-_\nu\spc \Bigr) .
\eea
Notice that there is now an explicit factor of $N$ in front of the last term.

We now proceed to apply the same large $N$ logic as in 1D. We write the perturbation series for fixed $e$ as a sum of `iterated melon' diagrams \cite{KitaevTalks,Polchinski:2016xgd, Maldacena:2016hyu}. The fermions then have a standard kinetic term and propagator. At the end, we integrate out $e$, which diagrammatically amounts to connecting all $\psi_+$ and $\psi_-$ lines by an $e$ propagator 
\bea
\la \spc  e_\mu^+({\rm x}_1) \, e_\nu^-({\rm x}_2)\spc \ra \is \frac 1 N\;  {\epsilon_{\mu\nu}}  \delta(x_{12}). 
\eea
Note that each Wick contraction $\contraction[.5ex]{}{e'' }{\  }{\ \ }{\; e^+}$ $\! e^-$ 
 produces a factor of $1/N$.  

A slight problem with the procedure just outlined, however, is that the $\psi$ propagators are singular at $e=0$, which is the point around which we wish to define the perturbation series. So whenever the $e$-line connects to a $\psi$ propagator, the $\psi$ propagator in fact collapses to a point. This is not surprising, since  we are in fact trying to write a perturbative expansion for an action (\ref{action}) without any quadratic term.

A more practical approach is to recast the model in terms of bosonic bi-local dynamical mean fields, given by the two-point function $G_\pm(\xx_1,\xx_2)$ and  self-energies $\Sigma^\pm({\rm x}_1,\xx_2)$.\footnote{Here and in the following we use the same symbols for the dynamical fields as for the on-shell solutions.}
After performing the disorder average and integrating out the fermions, one obtains the following effective action 
\bea
\label{dmfaction}
S/N\is  -\sum_{a=\pm} \log {\rm Pf} (\epsilon^{\mu\nu} e^a{\!\!}_\mu \partial_\nu - \Sigma^a) - \int \!  \epsilon^{\mu\nu} e^+_\mu e_\nu^-  \nonumber\\[-5mm]\\[-.1mm]
\nonumber
& &   \ +\, \frac{1}{2} \int\!\! \int \bigl(\Sigma^a G_a  - \frac{J^2}{q} (G_+)^q (G_-)^q\bigr)
\eea
This effective action looks quite similar to the dynamical mean field action of the 1D SYK model \cite{KitaevTalks, Sachdev:2015efa, Maldacena:2016hyu}. The key new features are the doubling of the number of fields and the presence of the frame variable $e^a$.

Since the action has an overall factor of $N$, the Schwinger-Dyson equations for $G_\pm$ and $\Sigma^\pm$ reduce in the large $N$ limit to the following saddle point equations
\bea
\label{sdo}
\Sigma^\pm(\xx_{12})  \is J^2 G_\pm(\xx_{12})^{q-1} G_\mp(\xx_{12})^q 
,\\[4.5mm]
\label{sdt}
\bigl( \epsilon^{\mu\nu} e^\pm{\!\!\!\!}_\mu \, \partial_\nu G_\pm \hspace{-2.5mm} & & \hspace{-5.5mm} -\;\, \Sigma^\pm *\spc  G_\pm\bigr)(\xx_{12}) \,=\, \delta^2(\xx_{12}),
\\[3.5mm]
\label{sdf}
e^\pm_\mu(\xx_1)\hspace{-6mm}  & &  = \,  { \partial  G_\mp(\xx_{12}) \over \partial \xx_2^\mu}\, \Big|_{\xx_2\to \xx_1} .
\ea
The $*$ in equation (\ref{sdt}) denotes the convolution product. Assuming translation symmetry, equation (\ref{sdf}) yields a constant value for $e^a_\mu$, which on
dimensional grounds, is proportional to the temperature $\beta^{-1}$ times~$(\beta^2 J)^{-1/q}$. 

We can represent the SD equations in diagrammatic notation as follows. Let us denote the chiral factorization equation (\ref{factorize}) as
\bea
{\cal F}(\xx_1,\xx_2) \is \; \; \raisebox{-23pt}{\includegraphics[scale=0.47]{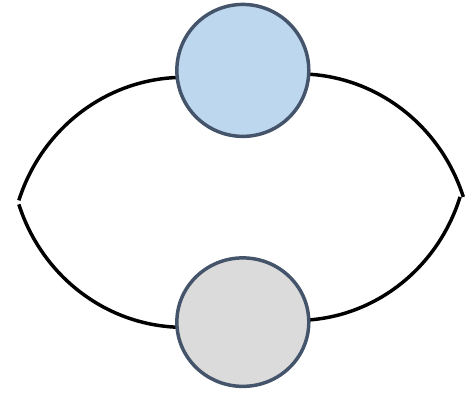}}
\ea

\noindent
where each line with a blob represents a dressed propagator $G_\pm(\xx_{12})$ of the chiral Majorana fermion. The color of the blob represents whether it is a $\psi_+$ (blue) or a $\psi_-$ (grey).

\begin{figure}[t]
\begin{center}
\includegraphics[scale=0.5]{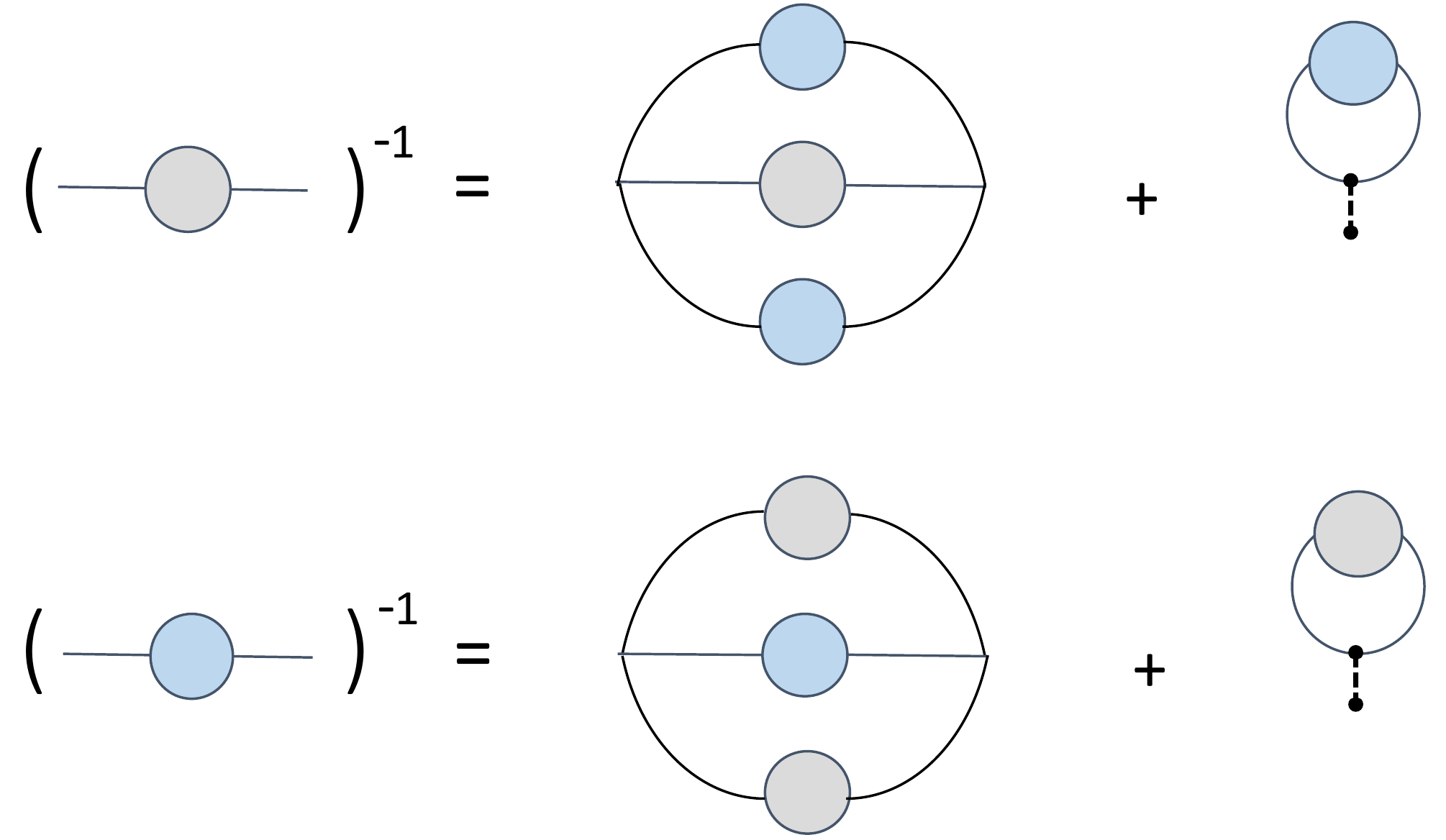}
\caption{\label{sdeqns} Diagrammatic representation of the SD equations \eqref{sdo}, \eqref{sdt} and \eqref{sdf} for $q=2$.}
\vspace{-2mm}
\end{center}
\end{figure}

The SD equations for $q=2$ are then depicted as in figure \ref{sdeqns}. The left-hand side denotes each inverse propagator, while first term on the right-hand side denotes the self-energy. The second term is the contribution from the dynamical kinetic term, which takes the form of a tadpole diagram attached via an $e$ propagator.

\subsection{Conformal limit}

\vspace{-1mm}

Continuing the standard SYK logic, we first focus on the IR limit. The interaction term then dominates and, working to leading order in $\beta^2 J$, we can drop the UV term. The SD equation (\ref{sdt}) then truncates to
\bea
\label{siginv}
\bigl(G_\pm * \Sigma^\pm\bigr)(\xx_{12})\, =\, -\delta(\xx_{12}).
\eea
In momentum space (and assuming translation invariance) this further simplifies to 
\bea
G_\pm({\rm k}) \Sigma^\pm({\rm k}) = -1.
\eea
We will solve  equations \eqref{sdo} and \eqref{siginv} via a scaling Ansatz momentarily.

Equations (\ref{sdo})-(\ref{siginv}) are diffeomorphism invariant, and as in the 1D model, this points to a zero mode of the linearized SD equations. In the following subsection, we will exhibit this zero mode by studying the four point function. As a quick preparation, consider a change in $G_\pm$ that corresponds to a reparametrization $(x^+,x^-) \to (x^+ + \epsilon^+,x^- + \epsilon^-).$ In the IR limit this is still a solution of SD equations if we take the Green function and self energy to transform accordingly. The variation of equation (\ref{siginv}) gives the  conditions
\bea
\delta G_\pm \ast \Sigma^\pm + G_\pm \ast \delta \Sigma^\pm \is 0.
\ea
We can take the product on the right by $(\Sigma^\pm)^{-1}=G_\pm $ to isolate $\delta G_\pm$ and use the expression for $\Sigma^\pm$ in terms of $G_\pm$ in the second term to eliminate the self energy from the equation. The above equation then takes the form, {\it c.f.} \cite{KitaevTalks}\cite{Maldacena:2016hyu}
\bea
(\delta_{ab}-K_{ab})\ast \delta_\epsilon G_b \is  0.
\eea
where $K_{ab}$ is the integration kernel 
\bea
\label{kernel}
K_{ab}\spc (x_1...\spc x_4)\! \is \!
-J^2 (q-\delta_{ab}) \spc 
G_a (x_{13})G_a (x_{24})\spc L_{ab}(x_{34})\nonumber \\[1mm] \\[-4mm]\nonumber
 L_{ab} \hspace{-8.5mm}&&(x) \, =\,
\frac{G_+(x)^{q}\, G_-(x)^{q}}{G_a(x)\spc G_b(x) } 
\ea
This shows that the eigenvalues of the kernel are $1$ when evaluated at reparametrizations of the conformal answer. Below we will make this formal conclusion explicit.

\def\sttrut{\raisebox{-.5pt}{\footnotesize${\strut}$}}
\def\sstrut{\raisebox{2pt}{\footnotesize${\strut}$}}

\subsubsection{Two-point function}

\vspace{-1mm}

We will now study the SD equations, following the approach of  \cite{ParcolletGeorges}. In the IR regime, we adopt the following scaling Ansatz for the dressed propagators and self energies
\bea
G_\pm\hspace{-5mm} & & \hspace{-4mm}\spc ({\xx})\;  = \,\spc b  \, \frac{{\rm sgn}(x^\pm)}{|x^+|^{\Delta\pm s}\spc |x^-|^{\Delta \spc \mp \spc s}}\nonumber\\[-1.5mm]
\label{ansatz}\\[-2mm] \nonumber
\Sigma_\pm({\xx})\! \is   J^2 \, b^{2q-1}\,  \frac{{\rm sgn}(x^\pm)}{|x^+|^{2-\Delta\spc \mp \spc s}\spc |x^-|^{2-\Delta\spc \pm \spc s}}
\ea 
with $b$ some constant. Here $\Delta$ and $s$ denote the sum  and difference of the left- and right scale dimensions. In the following, we will sometimes use the notation 
$\Delta_\pm = \Delta \pm s$. The sign functions in \eqref{ansatz} implement Fermi statistics, and match with 2-point function of the UV theory.  The IR Ansatz breaks the diffeomorphism invariance of the IR theory.  A new feature of the 2D model, relative to the 1D case, is that the sign and scaling functions specify a choice of light-cone direction and a signal propagation speed. 

The Ansatz \eqref{ansatz} solves the SD equations \eqref{sdo} and \eqref{siginv} provided that $\Delta = 1/q$ and
\bea
\label{norm}
J^2 b^{2q} \is \frac{((1-\Delta)^2 - s^2)}{4\pi^2 \cot\bigr(\frac \pi 2 (\Delta \! \spc +\!\spc s)\bigr) \tan\bigr(\frac \pi 2 (\Delta\! \spc -\! \spc s)\bigr)} \, \equiv \, \alpha^2_{sq}.\quad\ \
\ea
The value of the spin $s$ is not determined by the SD equations.\footnote{ A similar issue appears in \cite{Fu:2016vas} for the supersymmetric SYK model.}
For generality, we will treat $s$ as a free parameter. The most reasonable and consistent choice is to set $s=\Delta$. We will call this the chiral limit, as it preserves the property that $\psi_+$ and $\psi_-$ depend only on one light cone coordinate.  Note, however, that the $s \to \Delta$ limit has to be taken together with a $J\to \infty$ limit, while keeping $b$ fixed.  

\def\sttrut{\raisebox{-.5pt}{\footnotesize${\strut}$}}
\def\sstrut{\raisebox{2pt}{\footnotesize${\strut}$}}

\begin{figure}[t]
  \begin{center}
  \begin{tikzpicture}[scale=0.75]
  \node[] at (-4.5,1) {\small $K_{++}=$};
\draw [line width=0.5mm, blue] (0-1,2) arc (45:-45:1.41);
\draw [line width=0.5mm, gray] (0-1,2) arc (90+45:180+45:1.41);
\draw [line width=0.5mm, blue] (-2-1,0) -- (2-1,0);
\draw [line width=0.5mm, blue] (-2-1,2) -- (2-1,2);
\node[] at (1.15-1,1) {\footnotesize \textcolor{blue}{$q\! -\! 2$}};
\node[] at (-0.75-1,1) {\footnotesize $q$};
  \node[] at (3.5,1) {\footnotesize $K_{+-}=$};
\draw [line width=0.5mm, blue] (6+1,2) arc (45:-45:1.41);
\draw [line width=0.5mm, gray] (6+1,2) arc (90+45:180+45:1.41);
\draw [line width=0.5mm, blue] (4+1,2) -- (6+1,2);
\draw [line width=0.5mm, gray] (6+1,2) -- (8+1,2);
\draw [line width=0.5mm, blue] (4+1,0) -- (6+1,0);
\draw [line width=0.5mm, gray] (6+1,0) -- (8+1,0);
\node[] at (8-2.1,1) {\footnotesize$q\! -\!1$};
\node[] at (8.2,1) {\footnotesize \textcolor{blue}{$q\! -\!1$}};
\end{tikzpicture}

\bigskip
\bigskip

  \begin{tikzpicture}[scale=0.75]
  \node[] at (-4.5,1) {\small $K_{-+}=$};
\draw [line width=0.5mm, gray] (6+1,2) arc (45:-45:1.41);
\draw [line width=0.5mm, blue] (6+1,2) arc (90+45:180+45:1.41);
\draw [line width=0.5mm, gray] (4+1,0) -- (8+1,0);
\draw [line width=0.5mm, gray] (4+1,2) -- (8+1,2);
\node[] at (7+1.1,1) {\footnotesize$q\! -\! 2$};
\node[] at (7-0.75,1) {\footnotesize\textcolor{blue}{$q$}};
  \node[] at (3.65,1) {\small $K_{--}=$};
\draw [line width=0.5mm, gray] (0-1,2) arc (45:-45:1.41);
\draw [line width=0.5mm, blue] (0-1,2) arc (90+45:180+45:1.41);
\draw [line width=0.5mm, gray] (-2-1,2) -- (0-1,2);
\draw [line width=0.5mm, blue] (0-1,2) -- (2-1,2);
\draw [line width=0.5mm, gray] (-2-1,0) -- (0-1,0);
\draw [line width=0.5mm, blue] (0-1,0) -- (2-1,0);
\node[] at (1.15-1,1) {\footnotesize $q\! -\!1$};
\node[] at (-1.15-1,1) {\footnotesize\textcolor{blue}{$q\! -\!1$}};
\end{tikzpicture}
\end{center}
\caption{\small \it Diagrammatic definition of the kernel that gives the four-fermion correlation function. Here each line represents multiple dressed propagators, with  multiplicity as indicated.}
\label{figure:LL}
\end{figure}
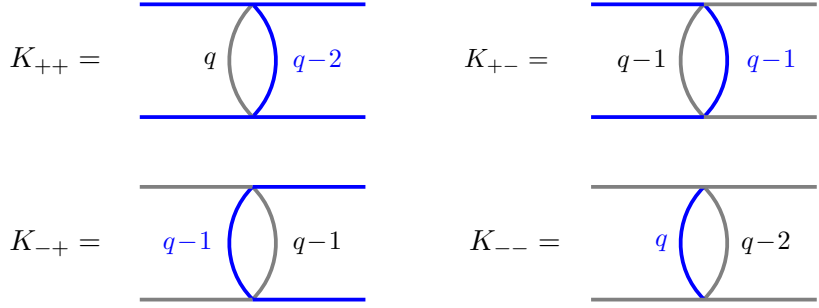

\subsubsection{Four-point function}
\def\KK{{\mathbb{K}}}
\vspace{-1mm}

Next we study the following four types of four-point functions
 \bea
{\cal F}_{ab}(x_1\, ...\, x_4) \is  \frac{1}{N^2} \sum_{i,j} \lb \psi_a^i(x_1) \psi_a^i (x_2) \psi_b^j(x_3) \psi_b^j(x_4)\rb \nonumber
 \ea
 with $a,b = \pm$. Like the two-point functions, these have to be thought of a part of a locally left-right symmetric correlation function.
The $1/N$ corrections to these four point functions can be computed with the help of the kernel $K_{ab}$  introduced in \eqref{kernel}. 
 
To leading order in $1/N$, we have
\bea
\mathcal{F}^{(0)}_{ab} = \bigl( - G_a(x_{13}) G_a (x_{24}) +  G_a(x_{14})G_a(x_{23}) \bigr)\delta_{ab} .
\eea
As explained in \cite{KitaevTalks,Polchinski:2016xgd, Maldacena:2016hyu}, the $1/N$ corrections to ${\cal F}_{ab}$ are found by summing up the $n$-th order contributions ${\cal F}^{(n)}_{ab}$ defined via the recursive formula
\bea
\mathcal{F}_{ab}^{(n+1)} \is \spc 
\sum_{c} K_{ac} *
 \mathcal{F}_{cb}^{(n)}
\eea
where $K_{ab}$ denotes the kernel \eqref{kernel} and $*$ denotes the double convolution product defined by identifying and integrating over the last two coordinates of $K_{ab}$ and the first two coordinates in  $\mathcal{F}_{cb}^{(n)}$. The diagrammatic form of the matrix elements of the kernel is depicted in figure \ref{figure:LL}.

The iterative procedure gives the following expression 
\bea
\label{geomsum}
\mathcal{F} \is \frac{1}{1- K* }\; \mathcal{F}^{(0)}.
\ea
where we have absorbed the matrix product into the definition of $*$. Inserting the conformal Ansatz \eqref{ansatz} into \eqref{kernel} gives the factorized expression
\bea
K_{ab} \is  - \frac{1}{\alpha_{ab}} K_{ab}^+(x^+_i) \spc K_{ab}^-(x^-_i)\\[1mm]
& &\hspace{-5mm} \frac{1}{\alpha_{ab}} \, = \, (q - \delta_{ab})\spc \alpha^2_{sq} 
\eea 
with $\alpha_{sq}$ defined in equation \eqref{norm}. 

Two representative examples of the chiral kernels are
\bea
K_{++}^- \is \,
\frac 1{  |x^-_{13}|^{\Delta_-} |x^-_{24}|^{\Delta_-} |x^-_{34}|^{2-2\Delta_-}\!\!\!\!\!} \\[2.5mm]
K^+_{+ +} \is \,
\frac{{\rm sgn}(x^+_{13})\, {\rm sgn}(x^+_{24}{\sttrut})}{\sstrut |x^+_{13}|^{\Delta_+} |x^+_{24}|^{\Delta_+} |x^+_{34}|^{2-2\Delta_+}}
\eea
with $\Delta_\pm = \Delta \pm s$.
The action of $K_{ab}^\pm$ on the four point functions can be computed with the standard SYK technique, by decomposing each ${\cal F}_{ab}$ in terms of eigenfunctions of the conformal Casimir  \cite{KitaevTalks,Polchinski:2016xgd,Maldacena:2016hyu}. These eigenfunctions are given by the three point functions of the fermion with an operator of some given left and right conformal dimension $(h, \bar{h})$.
A novel feature of our model  is that the eigenvalues of the kernels are given by two different types of integrals. One type of integral looks SYK-like $\int\! dx_1 \spc dx_2 \, K_{++}^+(0,1,x_1,x_2) {{\rm sgn}(x_{12})}{|x_{12}|^{h-\Delta_+}\!\!\!\!}$\;\;.  We denote the corresponding  eigenvalue by $k_{\Delta_+}(h)$. The other type of integral looks like $\int\! dx_1 dx_2 \, K_{++}^-(0,1,x_1,x_2) {|x_{12}|^{h-\Delta_-}}$. We denote the corresponding  eigenvalue by $\tilde{k}_{\Delta_-}(h)$. When acting on an eigenstate the kernel then takes the form
\bea
K_{ab} \is \frac{1}{\alpha_{ab}} \left(\begin{array}{cc} k_{\Delta_+}\!(h) \spc \tilde{k}_{\Delta_-}\!(\bar{h})\, &\,  \frac{q}{q-1} \tilde{k}_{\Delta_-}\!(h)\spc k_{\Delta_+}\!(\bar{h}) \\[2.5mm] \frac{q}{q-1} \tilde{k}_{\Delta_-}\!(h)\spc k_{\Delta_+}\!(\bar{h})\, &\,  \tilde{k}_{\Delta_-}\!(h)\spc k_{\Delta_+}\!(\bar{h})\end{array}\right)
\eea

The kernel $K_{ab}$ gives useful information about the spectrum. As a first consistency check, let us act with $K$ on an eigenmode with conformal dimension $(h, \bar{h}) = (2,0)$. This mode corresponds to the stress tensor, and is expected to describe the effective Goldstone mode associated with broken reparametrization invariance. We find that
\bea
K_{ab}{}_{\bigl\vert}{}_{\strut \mbox{\scriptsize$\begin{array}{c}\!\! h=2\\[-2mm] \!\! \bar{h}=0 \end{array}$}} \is  \left(\begin{array}{cc} \frac{(-1+\Delta) (\Delta+s)}{\Delta(2-\Delta-s)} \, &\, \frac{-\Delta -s}{\Delta(2-\Delta-s)} \\[1.5mm] \frac{-\Delta+s}{\Delta (2-\Delta+s)}\, &\,   \frac{(-1+\Delta)(\Delta-s)}{\Delta(2-\Delta+s)}\end{array}\right)
\eea
which manifestly satisfies $\det(\mathbb{1} - K) = 0$. Hence the intermediate states with scale dimension $(2,0)$ and $(0,2)$ appear as poles in the conformal strong coupling limit of the expression \eqref{geomsum}. 

\begin{figure}[t]
\begin{center}
\includegraphics[scale=1]{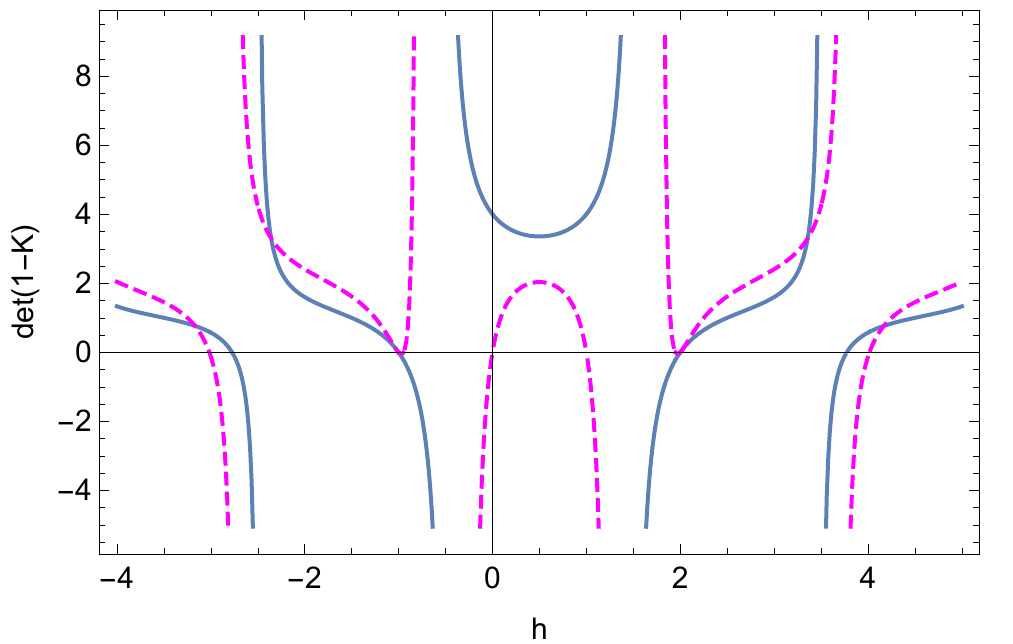}
\vspace{-3mm}
\caption{\small \it Plot of $\det(\mathbb{1} - K)$  as a function of the left scale dimension $h$ with $\bar{h}=0$. The dashed magenta plot corresponds to $\Delta = 1/4$ and $s=1/2$, and the blue plot to $\Delta = s = 1/4$. \label{fig:plot}}
\end{center}
\vspace{-4mm}
\end{figure}

In figure \ref{fig:plot} we have plotted $\det(\mathbb{1} - K)$ for chiral intermediate states with left scale dimension $h$ with $\bar{h}=0$. For illustration, we also included the case $s=1/2$ given by the dashed magenta graph. The blue graph corresponds to $s= \Delta = 1/q$, which is the chiral limit with $\Delta_- \! = 0$. In both cases we have set $q=4$. We see the expected symmetry between $h$ and $1-h$. For $s=1/2$, there are additional zeroes at $h=0$ and $1$, which indicates the possible presence of a spin one current, c.f. \cite{Fu:2016vas}. The spectrum of zeroes of the $s=\Delta =1/4$ case, on the other hand, looks identical to that of the SYK model. This theory is a plausible candidate for a 2D QFT with maximal chaos. The explicit formula for $\det(\mathbb{1}-K)$ for $\bar{h}=0$, $s=\Delta = 1/q$ is
\bea
\det(\mathbb{1}-K)\, =\; 1 \, +\, 
\frac{\pi ^2 (q-2) (q-1) \csc(\frac{2 \pi }{q})}{q\spc \Gamma(\frac{2}{q})^2 \bigl(\sin (\pi  h)\!\spc +\!\spc \sin(\frac{2 \pi }{q})\bigr) \Gamma(2\!\spc-\!\spc \frac{2}{q}\!\spc -\!\spc h) \Gamma(1\!\spc-\!\spc\frac{2}{q}\!\spc +\!\spc h)}\nonumber
\ea
which coincides with the expression for $1-K$ in the SYK model \cite{KitaevTalks,Polchinski:2016xgd,Maldacena:2016hyu}.
We leave a detailed calculation of the four-point function and the spectrum of states for future work. 

\vspace{-1mm}
\section{Effective action of the Goldstone mode}\label{sec:EffectiveAction}
\vspace{-1mm}

We would like to exhibit the effective action of the reparametrization mode. In principle, we could try to follow the procedure used in \cite{KitaevTalks,Maldacena:2016hyu}, compute the correction to the kernel $K_{ab}$ that follows from including the UV term of the action \eqref{action}, and use this to find the linearized action of the zero modes. We reserve this calculation for a future project. Here we will instead make a short-cut, which appears justified in the case that $q$ is small enough so that an expansion in $\epsilon =1-2/q$ is valid \cite{Jevicki:2016ito}. Note that in our model, the $q=2$ system is still an interacting QFT.

\subsection{Double Schwarzian action}
\vspace{-1mm}

We start from the dynamical mean field action \eqref{dmfaction},  and perform the redefinition $\Sigma^\pm_{new} =   \Sigma^\pm_{\rm old} -\epsilon^{\mu\nu}e^\pm_\mu \partial_\nu.$ This redefinition moves all the $e^\pm$ dependence into a separate UV term
\bea
& & \qquad\qquad \  S/N\, = \, S_{\rm UV} + S_{\rm IR}\nonumber\\[4.5mm]
\label{seff} 
S_{\rm IR} \is  -\, \sum_{a=\pm} \, \log {\rm Pf} ( \Sigma^a) \, +\, \mbox{\Large $\frac{1}{2}$} \int\!\! \int \Bigl(\Sigma^a G_a  - \mbox{\Large $\frac{1}{q}$} J^2 \, (G_+)^q (G_-)^q\Bigr)\\[2mm] 
& & \quad  S_{\rm UV} \, = \,  \frac 1 2  \int \!  d^2\xx \, \epsilon^{\mu\nu}\epsilon_{ab} \, \bigl( e^a_\mu G^b_\nu - e^a_\mu e^b_\nu\bigr)
\label{uvcorr}
\eea
Here we defined 
\bea
G_\mu^a(\xx_1) \is \epsilon^{ab} \, {\partial G_b (\xx_{12}) \over \partial \xx_2^\mu}\Big|_{\xx_2\to \xx_1}
\eea
The IR term is the same as before, and leads to the conformal and reparametrization invariant equations of motion \eqref{sdo} and \eqref{siginv}. However, because 
$\Sigma^a$ is the shifted variable, equation \eqref{siginv} is now exact, and equation \eqref{sdo} receives a subleading correction due to the presence of the UV term \eqref{uvcorr}. An exact treatment of the consequences of this correction term could be accessible in the large $q$ limit \cite{Maldacena:2016hyu}. We will instead look at the regime  $q = 2/(1 -\epsilon)$ with $\epsilon$ small, and restrict our attention to the chiral limit $s=1/q$.

The total bosonic action \eqref{seff} has the same invariances as the original fermionic action \eqref{action}, namely  (i) area preserving diffeomorphisms, and (ii) local Lorentz transformations. These symmetries are shared by the UV and IR terms in the action and we will treat both as gauge symmetries. The IR action is also invariant under local conformal transformations\footnote{In the rest of this subsection, we temporarily move the upper $\pm$ index on $x^\pm$ to a lower index.}
\bea
\label{conf}
(u,v) \to (x_+(u),x_-(v)).
\eea
This conformal symmetry is broken in two separate ways.
Picking a particular conformal IR solution of the SD equations spontaneously breaks the local conformal invariance to the global conformal group. This leads to the presence of a Goldstone mode, parametrized by the conformal transformation $(x_+(u), x_-(v))$. Moreover, the UV action is not invariant under the same local conformal transformation rule as the IR action. So it induces a non-trivial effective action for the Goldstone mode.

To get the leading order form of the effective action of the Goldstone mode, we perform a local conformal transformation on the IR propagator. It transforms as
\bea
\label{trafo}
 G_\pm(u,v,\tilde{u},\tilde{v}) \is  [x'_+ x_+' ]^{\Delta_\pp} [x_-' \tilde{x}_-']^{\Delta_\mm}\; G_\pm\bigl(x_\pm, \tilde{x}_\pm\bigr)\qquad
\eea
with $(x_\pm,\tilde{x}_\pm)  = \bigl(x_\pp(u),x_\mm(v), x_\pp(\tilde{u}), x_\mm(\tilde{v})\bigr)$ and $x_+' = \partial_u x_+(u)$, etc.
In the chiral limit $\Delta_- \!\! \to 0$ and $\Delta_+\!\!\spc =\spc \! \Delta\!\spc =\spc \!  1/q = \frac 1 2 (1- \epsilon)$, the dressed propagator behaves in the conformal regime as $G_\pm(x,\tilde{x}) = \frac{b^{1-\epsilon}}{|x_\pm - \tilde{x}_\pm|^{1-\epsilon}}$. We can now use this expression, transform it via \eqref{trafo}, plug it into the UV action \eqref{uvcorr}, and extract the dependence on $x_\pp(u)$ and $x_\mm(v)$. 

The conformal propagator diverges in the coincident limit. This divergence is expected to be removed by the UV modification. A more practical method is to take the coincident limit while subtracting the singular contribution in the $(u,v)$ coordinates. Working to leading order in $\epsilon$ and using that
\bea
\partial_u \left( \frac{\sttrut \sqrt{x'(u) x'(\tilde{u})}}{\sstrut |x(u)-x(\tilde{u})|} - \frac{\sttrut 1}{\sstrut |u-\tilde{u}|}\right){\!}_{\Bigl\vert_{\strut{{}^{\mbox{\small ${\tilde{u}=u}$}}}}}\hspace{-2mm}
 \is \frac{1}{12} 
\bigl\{x,u\bigr\} \ \ 
\eea
with 
$
\{\spc x,u\spc \}
=\frac{x{'''}}{{x'}} -\! \frac 3 2 \bigl(\frac{x{''}}{{x'}}\bigr)^2
$
 the Schwarzian derivative,
we find that the effective action of the reparametrization modes 
takes the form
\bea
\label{sumschwarz}
\frac {S_{\rm UV}} N \is  \frac{b}{12}  \int\! du \spc dv \, \bigl(e^+_v \bigl\{x_\pp,u \bigr\} \spc + \spc e^-_u \bigl\{x_\mm,v\bigr\}\bigr) - \int \epsilon^{\mu\nu} e^+_\mu e^-_\nu
\eea
After integrating out $e^\pm$, we obtain 
\bea
\label{doubleschwarz}
\frac {S_{\rm UV}} N \spc \is\,  \frac {\alpha_S}{J} \spc \int\! du\spc dv \, 
\bigl\{x_\pp,u \bigr\} \spc \bigl\{x_\mm,v\bigr\}.
\eea
Applying the $\epsilon$ expansion method of \cite{Jevicki:2016ito} gives that
\bea
\alpha_S \is \frac{\alpha_{sq}}{144}\, \bigl(1 -  2\epsilon^2) \, +\,  O(\epsilon^4)
\eea
with $\alpha_{sq}$ defined in \eqref{norm}. 
The effective action \eqref{doubleschwarz} is a functional on the group of the 2D conformal transformations. It generalizes the Schwarzian action for the reparametrization mode of the SYK model. We expect that, by generalizing the analytic and numerical analysis of \cite{Maldacena:2016hyu} to the 2D model, it should be possible to compute the pre-coefficient $\alpha_S$ for general values of $q$. 

\def\TT{\mbox{\small $T$}}

\subsection{Free energy and spectral density} 

\vspace{-1mm}

By considering the transformation of the Schwarzian derivative under conformal mappings, we can extract useful information about the behavior of the theory on a circle and at finite temperature. At finite temperature, the effective action \eqref{doubleschwarz} receives additional terms 
\bea
\label{tcorrect}
\frac{\alpha_S} {J}\! \int\!\! \!\! &du\spc dv&\!\!\!\! \spc \Bigl(\TT_{--} \{x^+\!\spc,u\}  +  \TT_{++} \{x^-\!\spc, v \} \spc +  \spc T_{++}T_{--} \Bigr)\ \ 
\nonumber\\[1mm]
\TT_{++}\hspace{-10mm} & &\hspace{-5mm} = \, \frac{\pi^2}{\beta_+^2} (x'_+)^2, \qquad\; \TT_{--} = \, \frac{\pi^2}{\beta_-^2} (x'_-)^2,
\eea
with $\beta_\pm$  the left- and right-moving inverse temperature. 
This term is subdominant at low temperature, but becomes important at distance scales of order the thermal wave length. If we take, say, the left moving high temperature limit,  we obtain a single Schwarzian action for the right-movers. This suggests that the 2D model reduces to the 1D SYK model by performing a DLCQ limit.

Equation \eqref{doubleschwarz} captures the explicit breaking of conformal invariance of the IR theory due to UV term in  \eqref{action}. Its form as a product of two chiral Schwarzian derivatives, as well as the finite temperature correction term \eqref{tcorrect}, indicates that the leading order correction to the IR conformal field theory takes the form of an irrelevant
$T\bar{T}$ deformation, given by the product of the left- and right-moving stress tensors \cite{nyu,smirzam,italians, ttbar}.

 To test this interpretation, let us consider the model on a cylinder with circumference $L=2\pi$.\footnote{So all dimensionful quantities are measured in units of the cylinder radius.} The conformal mapping from the plane to the cylinder induces a negative Casimir energy, which can be taken into account by setting $\{x_\pp,u\}\! = \{x_\mm,v\} =\! - 1/2$ in equations~\eqref{doubleschwarz}-\eqref{tcorrect}. Now consider the contribution of the effective action of the Goldstone mode to the free energy at finite temperature. Setting $\beta_\pm = \beta$, we find that
\bea
\label{supset}
-\beta F & \supset & -\frac { N \alpha_S}{J}  \int_0^{2\pi}\!\!\!\!\! dx \int_0^\beta \!\!\! dt  \, \Bigl(- \frac 1 2 + \frac{\pi^2}{\beta^2}\Bigr)^2 \nonumber\\[-2mm]\\[-2mm]
\is  \frac { 2\pi N \alpha_S}{4J} \Bigl(-\beta  \spc + \spc \frac{4\pi^2}{\beta}\spc - \spc \frac{4\pi^4}{\beta^3} 
\spc \Bigr)\, .\nonumber
\eea

We wish to compare this result with the free energy of a CFT of central charge $c$ with a $T \bar T$ deformation. The energy spectrum and thermodynamics of this class of theories was studied in detail in \cite{nyu,smirzam,italians}, and a holographic interpretation\footnote{The holographic dual of the $T \bar{T}$ deformation proposed in \cite{ttbar} is closely similar to the candidate AdS${}_2$ dual interpretation of the 1D SYK model developed in \cite{Jensen:2016pah,Maldacena:2016upp,Engelsoy:2016xyb}, built on the earlier work  \cite{AP}. In both cases, the boundary of the AdS space-time is moved into the bulk. On the CFT side, this represents an explicit breaking on conformal invariance and gives rise to an associated dynamical pseudo-Goldstone mode. We will make the AdS${}_3$ interpretation of the action \eqref{doubleschwarz} more explicit in the next subsection.} 
   has recently been proposed in \cite{ttbar}. 
Using the results of \cite{nyu}\cite{ttbar},  one finds that the free energy of a deformed CFT with action $S_{\rm CFT} + \int \mu\, T \bar T$ has the following small temperature expansion
\bea
\label{ttbar}
-\beta F_{{}{\rm{CFT \mbox{\tiny +} \mu T \bar T}} } \, \is\, -  \frac{c \spc \beta}{12}  \,  + \,  \frac{\pi^2 c}{3 \beta} \, -\, \frac{\pi^3 \mu c^2 }{72\beta^3} \, + \, ...
\eea
The first two terms are the standard CFT expression for the Casimir energy and specific heat.
Comparing the expressions \eqref{supset} and \eqref{ttbar} suggests that the IR limit of our model is a 2D CFT with central charge $c$,
and that the leading deviation on conformal invariance is given by a $T \bar T$ interaction with coupling $\mu$, with $c$ and $\mu$ given by
\bea
\label{cmu}
\frac c {24\pi} \spc \is \spc \frac {N \alpha_S}{4J}, \qquad \ \ \mu = \frac{24\pi}{c} = \frac{4J}{N \alpha_S}.
\eea
This reciprocal relation between $c$ and $\mu$ precisely agrees with the relationship derived from the holographic dictionary proposed in \cite{ttbar}.

By performing an inverse Laplace transform of the partition function $Z(\beta) = e^{-\beta F}$ with respect to $\beta$, we can extract the spectral density as a function of the energy~$E$:
\bea
\rho(E)&\! \propto \! & \exp\left(2\pi \sqrt{\frac{c E}{3}}\, \Bigl(1\spc - \spc  \frac{3 E}{2c} \spc + \spc ...\spc \Bigl) \right)
\eea
The leading term is the Cardy formula\footnote{Here $E$ is defined such that the CFT ground state has negative Casimir energy $E/L= -{c}/{12} $.} and the subleading term reflects the explicit breaking of conformal symmetry. 
This formula precisely matches with the low energy expansion of the exact equation of state  $E L - \frac \mu 4 E^2 = \frac 3 {2\pi c} S^2$  
relating the energy and the entropy $S=\log\rho(E)$ 
of the $T \bar T$ deformed CFT \cite{ttbar}, provided we set $L=2\pi$ and $\mu$ as in \eqref{cmu}.

\subsection{Relation with AdS${}_3$ gravity}\label{sec:EAgravity}

\vspace{-1mm}

The double Schwarzian action \eqref{doubleschwarz} can be related to the 3D AdS gravity action as follows. 

In the above derivation we identified the effective Goldstone  degree of freedom with the group of `passive' conformal reparametrizations \eqref{conf}. To match with the gravity side, it is convenient to represent the Goldstone mode as an `active' 2D conformal  transformation
\bea
\label{cinv}
(x^+,x^-) \to \bigl(\UU(x^+), \VV(x^-)\bigr)
\eea
defined as the inverse mapping of \eqref{conf}. In terms of ($\UU,\VV$), the effective action \eqref{doubleschwarz} reads
 \bea
 \label{seffnew}
 \frac{S[\UU,\VV]} {\alpha_S N /J}\is  \int\! d^2\xx \, \spc {\cal S}_+(\UU) \spc {\cal S}_-(\VV), 
 \eea
 where ${\cal S}_+(\UU)$ and ${\cal S}_-(\VV)$ are defined via
 \bea
{\cal S}_+(\UU)\, \partial_+\!\spc \UU \! \is\!  \{ \UU,x^+\!\spc\},  \qquad \quad
{\cal S}_-(\VV)\, \partial_-\!\spc \VV  \, = \,   \{ \VV,x^-\! \spc \}.
 \eea
 
We will now show that the effective action \eqref{seffnew} is equal to the 3D gravity action
\bea
\label{hologram}
S[\UU,\VV] \is  \, S_{\rm grav}[\UU,\VV]
\eea
evaluated on a suitable classical solution of 3D gravity defined on a AdS${}_3$ space-time with finite radial cut-off, specified as follows. Let $B$ denote the boundary of the cut-off AdS$_3$ space-time. We define the Einstein action via
\bea
S_{\rm grav} \is \frac{1}{16\pi G_N} \int\!\! \sqrt{g}\bigl(R - 2\Lambda\bigr) + \frac{1}{8\pi G_N} \int_B (K + 1)\nonumber
\eea
where, besides the usual extrinsic curvature term $K$, we included a boundary cosmological constant identical to the standard counter term used in holographic renormalization.
The classical solution associated with $(\UU,\VV)$ is defined via the boundary condition that the pull back of the 3D bulk metric to $B$ is a flat 2D metric given by 
\bea
ds^2{\vert_B} = d\UU d\VV \, = \, \UU'(x^+)\VV'(x^-)dx^+dx^-.
\eea 
The holographic identification \eqref{hologram} holds if we identify the bulk Newton constant as
\bea
\frac 1 {16\pi G_N} = \frac {N\alpha_S} {4J}.
\eea

Equation \eqref{hologram} looks a little surprising at first. One might think that, since the gravity action is reparametrization invariant, it should be independent of $\UU(x^+)$ and $\VV(x^-)$. Recall, however, that the Lagrangian changes by a total derivative under an active diffeomorphism, and that 2D conformal transformations necessarily extend all the way to null infinity. A helpful way to visualize the asymptotic region is by mapping the 2D space-time onto a Penrose diagram. The conformal transformations are then analogous to the BMS group. Once we choose a preferred reference coordinate system, the dependence of the action on  $\UU(x^+)$ and $\VV(x^-)$ becomes finite and computable. 

The holographic identification \eqref{hologram} can be derived in various ways. One is direct computation. Another route is to show that the action \eqref{seffnew}  satisfies the Hamilton-Jacobi equation that governs the radial evolution of a classical action in 3D gravity.
An instructive derivation goes via the following three basic steps. 

First we reintroduce 
the frame variables $e^\pm$ and rewrite \eqref{seffnew} as the minimum over $e^\pm$~of
\bea
\int \!\! d^2 x \spc \Bigl( e_-^- \spc {\cal S}_+\!\spc(\UU)\! + e^+_+\spc {\cal S}_-\!\spc(\VV)\! -\!\spc  \epsilon^{\mu\nu} e_\mu^+ e_\nu^-  \Bigr)  \label{hstwo}
\eea
Next we introduce the background zweibein 
\bea
\label{euv}
E^+_\mu dx^\mu \is d\UU, \qquad  E^-_\mu dx^\mu = d\VV
\eea
and make use of the relationship between the Polyakov-Liouville action (viewed as a functional of the zweibein)
\bea
S_\text{\spc L}[\spc E\spc ] \is  {1 \over 8\pi}\! \int \! R \,\square^{-1} R , \qquad \quad
 g_{\mu\nu} \spc =  \spc \eta_{ab}\spc E^a_{\mu} E_{\nu}^b
\eea
 and the Schwarzian derivative to write 
 \bea
 \label{mine}
\frac{S[\UU,\VV]}{\alpha_S N /J} \is \raisebox{-6pt}{$\raisebox{2pt}{min} \atop  \raisebox{1pt}{$e$}$}  \Bigl( 
 S_\text{\spc L}\bigl(E+ e\bigr) 
 \, -\,  \int \! \epsilon^{\mu\nu} e_\mu^+ e_\nu^- \Bigr).
\eea
Here we used that, in the linearized approximation, $S_\text{\spc L}[E + e] =  \int ( e^-_- \, {\cal S}_+(\UU) \spc +\spc e^+_+\,{\cal S}_-(\VV)).$
Note that the Polyakov action vanishes for the flat metric \eqref{euv} and that $\{\UU,x\} = -\frac 1 2 (\phi')^2 + \phi''$ with $\phi= \log \UU'$. 

Since the Polyakov action arises by integrating out a 2D CFT, the identity \eqref{mine} is yet another indication that the IR theory describes a 2D CFT in a fluctuating metric $g_{\mu\nu} = \eta_{ab}(E^a_\mu + e^a_\mu)(E^b_\nu + e^b_\nu)$. Integrating out the metric fluctuations first produces a  CFT with a $T \bar T$ deformation.

The third and final step in the derivation of \eqref{hologram}  uses an (underappreciated) result of Freidel that establishes a direct transformation between the 3D Einstein action evaluated on a classical background and the Polyakov action evaluated on the boundary metric \cite{freidel}
\bea
\label{freidel}
 S_\text{grav}[\spc E\spc ] \is  \raisebox{-6pt}{$\raisebox{2pt}{min} \atop \raisebox{1pt}{$e$}$} \Bigl( S_\text{\spc L}\bigl(E + e\bigr) 
 \, -\, \int \! \epsilon^{\mu\nu} e_\mu^+ e_\nu^-\Bigr) 
 \eea
Here $S_\text{grav}(E)$ is the classical bulk gravity action with boundary conditions $g_{\mu\nu} = \eta_{ab} \spc E^a_\mu E^b_\nu $. 
 The formula \eqref{freidel} forms the basis of the holographic interpretation of the $T \bar T$  deformed theory proposed in \cite{ttbar}. In our context, it provides the link between 3D gravity and the choice of kinetic term in our proposed 2D analog of the SYK model. A detailed derivation of the relation \eqref{freidel} can be found in \cite{freidel}.

\section{Conclusion}\label{sec:Conclusion}

\vspace{-1mm}

We have proposed a 2D QFT generalization of the SYK model, consisting of $N$ Majorana fermions with a random non-linear interaction. While the quartic kinetic term of our action \eqref{action} looks somewhat unconventional, it can be rewritten as in \eqref{hsaction} as a conventional quadratic kinetic term coupled to a dynamical metric. The total action is invariant under area preserving diffeomorphisms and local Lorentz transformations. We treat both invariances as gauge symmetries. 
 
 We have presented evidence that the model exhibits conformal symmetry in the IR, and that the low energy dynamics is dominated by  an emergent Goldstone-like mode associated with the breaking of conformal reparametrization symmetry.  Just as in SYK, this symmetry breaking is introduced by the fact that UV  action assigns a lower  scale dimension $[\psi]_{\rm UV} = 0$ to the Majorana fermions than the relevant interaction term, which prescribes that $[\psi]_{\rm IR} = 1/q$.  Some questions that need further study are:  Is there a principle that fixes the IR value of the spin $s$, or is it an adjustable parameter?  What do the Hilbert space, energy spectrum, partition function and correlation functions look like? 
 
 The motivation for our study is to find new examples of strongly coupled 2D QFTs with potential gravity duals and to elucidate the role of the reparametrization mode in the holographic dictionary. While our model still needs to be put on firmer footing, there are encouraging signs that it is well defined and exhibits the hallmarks of a holographic dual to AdS${}_3$ gravity. In particular, it seems plausible that the conformal symmetry is non-linearly realized in terms of the reparametrization mode. In our previous paper \cite{Turiaci:2016cvo} we have shown that this uniquely dictates the commutation relations of the Goldstone modes  and implies maximal Lyapunov growth of out-of-time ordered correlation functions. In view of the results of \cite{Turiaci:2016cvo}\cite{Jackson:2014nla} and the discussion in section \ref{sec:EAgravity}, we expect that the effective theory of the reparametrization mode should be closely related to Liouville theory. A natural route towards making this relationship concrete is to postpone the integral over Hubbard-Stratonovich variable $e^\pm$ and to extract its effective action by making use of equation \eqref{mine}. 
  
Finally, it is natural to speculate whether a similar approach could lead to proposed generalizations of the SYK model to higher dimensions. The UV action has an obvious reparametrization invariant generalization 
\bea
\epsilon^{ab...f}  \epsilon^{\mu\nu ...\sigma} \spc \psi_a \partial_\mu \psi_a \spc \psi_b \partial_\nu \psi_b\, ... \, \psi_f \partial_\sigma \psi_f. &&
\nonumber
\ea
Adding a $\psi^{q D}$ interaction term would again be a relevant deformation, and the two terms combined would be invariant under volume preserving diffeomorphisms. However, it seems premature to pursue this generalization without first obtaining a better understanding of the landscape of lower dimensional examples.

\bigskip

\begin{center}
{\bf Acknowledgements}
\end{center}

We thank Duncan Haldane,  Daniel Harlow, Kristan Jensen, Alexei Kitaev, Igor Klebanov, Juan Maldacena, Thomas Mertens, Mark Mezei, Mukund Rangamani, Douglas Stanford, Grisha Tarnopolsky and Zhenbin Yang for valuable discussions and helpful comments. The research of H.V. is supported by NSF grant PHY-1620059.

\bigskip

\begin{appendix}

\def\WW{\mbox{\small W}}
\section{Topological RCFT}\label{sec:App}

\vspace{0mm}

What is a topological RCFT? A rational CFT is a CFT with an infinite chiral algebra $\hat{\mathfrak{g}} \supset$ Vir  and a finite set of primary fields ${\cal O}_i$.  For minimal models, $\hat{\mathfrak{g}}$ equals the Virasoro algebra. A topological RCFT is defined by gauging the chiral algebra $\hat{\mathfrak{g}}$. This projects the operator content to the set of primary fields, and removes all position dependence of the Euclidean correlation functions of local operators. All known RCFT can be represented as coset WZW models, and all known topological RCFTs can be formulated as fully gauged WZW models. The topological Ising model is a gauged $\mathfrak{su}(2)_k$ coset with $k=2$. 

In Euclidean space, the four-point correlation function of local operators ${\cal O}_i = {\cal O}_i(\xx_i)$ in a TCFT are specified via the following simple rule \cite{HV,WZW-gauged, BlauThompson}
\bea
\la {\cal O}_{1} {\cal O}_{2}{\cal O}_3 {\cal O}_4\ra_{{}_{\rm TCFT}} = {\rm dim}\bigl( {\cal H}_{1234}\bigr), \nonumber
\eea
where ${\cal H}_{1234}$ denotes the linear vector space spanned by the chiral conformal blocks
\bea
{\cal F}_{\rm a}(1234)
 \is \, \raisebox{-6pt}{$ \includegraphics[scale=.4]{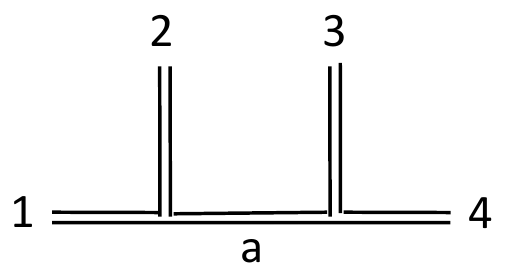}$}\, \nonumber
\eea 
associated with the corresponding CFT correlation function.
This rule satisfies all axioms of 2D TQFT \cite{robbert}. For gauged WZW models, the above prescription naturally follows from the identification of ${\cal H}_{1234}$ with the Hilbert space of a 3D Chern-Simons theory in the presence of four Wilson lines. Schematically
\bea
{\cal F}_{\rm a}(1234) \is \WW_1(1) \WW_2(2) \WW_3(3) \WW_4(4)  | 0 \rangle_{{\rm CS}}.
\eea
The CS functional integral on $\mathbb{R}^2 \times S^1$ reduces to the TCFT amplitude on $\mathbb{R}^2$ and takes the form of a sum of inner products between the left and right chiral conformal blocks \cite{HV,WZW-gauged, BlauThompson}
\bea
\la {\cal O}_{1}  {\cal O}_{2} {\cal O}_3 {\cal O}_4\ra_{{ \!\spc}_{\rm TCFT}} = \sum_{\rm a}
\; \langle {\cal F}_{\rm a} | {\cal F}_{\rm a} \rangle  =\, \tr_{{\cal H}_{1234}}\!\spc\bigl(\mathbb 1\bigr). \nonumber
\ea
Here we used the conventional RCFT normalization of conformal blocks, for which the fusion and braid operations are represented as unitary matrices. In this  unitary basis, the OPE coefficients are all given by integer fusion coefficients $N_{ijk}$. The local observables thus form a commutative, associative ring isomorphic to the fusion algebra 
\bea
{\cal O}_i \times {\cal O}_j = \sum_k N_{ijk} \, {\cal O}_k.
\eea

In Minkowski space-time, operator ordering plays a non-trivial role.  Local operators in an RCFT decompose as a sum of factorized terms ${\cal O}(x^+\! ,x^-) = \sum_s {\cal V}_s^+(x^+) {\cal  V}_s^-(x^-)$. The ${\cal V}^\pm_s$ are known as chiral vertex operators. Chiral vertex operators of the same chirality satisfy non-trivial braiding relations, and can be thought of as end points of light-like Wilson lines. Whenever a ${\cal V}_j$ passes through the light-cone of another ${\cal V}_{k}$, the corresponding chiral conformal block undergoes a non-trivial monodromy. E.g.
\bea
{\cal F}_{\rm a}(1234) \is \sum_{\rm b} R^{\spc \epsilon}_{\rm ab} {\cal F}_{\rm b}(1324) 
\qquad{\rm or} \qquad
\raisebox{-12pt}{$ \includegraphics[scale=.4]{ablock.pdf}$}\, =\, \sum_{\rm b} R^{\spc \epsilon}_{\rm ab} \;\; \raisebox{-12pt}{$ \includegraphics[scale=.4]{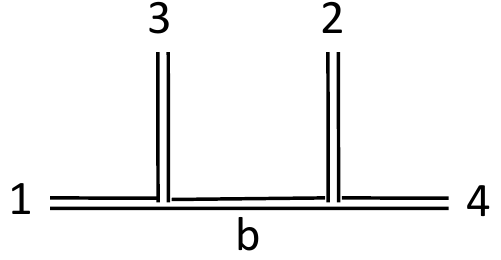}$}  \nonumber
\eea 
Here $R_{ab}$ is known as the R-matrix and $\epsilon =\pm 1$. This choice of sign indicates that the braiding move depends on orientation.  
The ordering of chiral vertex operators is encoded via the end-point of the corresponding Wilson lines, as indicated in figure \ref{wfactor}.

 In non-chiral correlation functions, the above monodromy produces a discontinuity when operators pass through each others light-cone. The monodromy of the left- and right-light cone have opposite orientation, so the total monodromy cancels out when two operators pass through both of each other's light-cones. Hence the Euclidean correlation functions are single valued and the non-chiral CFT thus remains local.

${}$~~The same chiral decomposition and dependence on operator ordering holds true in a topological RCFT.
The TCFT correlation functions thus acquire a non-trivial position dependence
\bea
\la {\cal O}_{1} {\cal O}_2  {\cal O}_3 {\cal O}_4\ra_{{ \!\spc}_{\rm TCFT}}\!  \is \, \sum_{{\rm a,b}}
\; \langle {\cal F}_{\rm a} | R^{\spc \epsilon}_{\rm ab} |{\cal F}_{\rm b} \rangle  =\, \tr_{{\cal H}_{1234}}\!\spc(R\spc) \nonumber
\eea
where $R^{\spc \epsilon}_{ab}$ is the $R$-matrix that implements the braiding operation that re-arranges all operators into space-like separated positions. 
The above discussion easily generalizes to higher $n$-point functions. 

Applying this general prescription to the special case of the $n$ point functions of the operators $\beps(\xx) = \psi_+(\xx)\psi_-(\xx)$ gives the result \eqref{npt}. The relevant $R$ matrix in this case is simply equal to the $(-1)$ factor that implements Fermi statistics.
The equality between \eqref{npt} and \eqref{nptpf} then follows directly by applying the definition of the Pfaffian
\bea
{\rm Pf}(M) = \frac{1}{2^n n!} \varepsilon_{i_1 j_1 i_2 j_2 ... i_n j_n} M_{i_1 j_1}  M_{i_2 j_2}... M_{i_n j_n}
\ea
for the case that $M_{ij} = {\rm sgn}(x_{ij})$ (with $x$ = $x^\pm$).
One can make a permutation of the $n$ points such that they are in order of increasing $x$. This can be recast as a permutation of the indices that gives an overall factor of $(-1)^P$ where $P$ is the parity of the permutation. Then the value is fixed by the Pfaffian when the points are ordered such that $x_1<x_2<...<x_n.$ This gives a factor of $(-1)^{n/2}$. Since $n$ is even, the total factor is equal to 
1  for the product \eqref{nptpf} of the left- and right-Pfaffian. 

\end{appendix}


\begin{thebibliography}{99}

\bibitem{KitaevTalks}
A. Kitaev, Talk given at the Fundamental Physics Prize Symposium, \href{https://www.youtube.com/watch?v=OQ9qN8j7EZI}{Nov. 10, 2014}; A. Kitaev, KITP
seminar, \href{http://online.kitp.ucsb.edu/online/joint98/kitaev/}{Feb. 12, 2015}; ``A simple model of quantum holography,'' talks at KITP,  \href{http://online.kitp.ucsb.edu/online/entangled15/kitaev/}{April 7, 2015} and \href{http://online.kitp.ucsb.edu/online/entangled15/kitaev2/}{May 27, 2015}.
  
  \bibitem{Sachdev:2015efa}
S.~Sachdev, ``{Bekenstein-Hawking Entropy and Strange Metals},''
  \href{http://dx.doi.org/10.1103/PhysRevX.5.041025}{Phys. Rev.
  {\bfseries X5} no.~4, (2015) 041025},
\href{http://arxiv.org/abs/1506.05111}{{\ttfamily arXiv:1506.05111 [hep-th]}}.

\bibitem{Polchinski:2016xgd} 
  J.~Polchinski and V.~Rosenhaus,
  ``The Spectrum in the Sachdev-Ye-Kitaev Model,''
  \href{http://link.springer.com/article/10.1007%2FJHEP04%282016%29001}{JHEP {\bf 1604}, 001 (2016)}
 \href{http://arxiv.org/abs/1601.06768}{{\ttfamily arXiv:1601.06768 [hep-th]}}.

\bibitem{Maldacena:2016hyu} 
  J.~Maldacena and D.~Stanford,
  ``Remarks on the Sachdev-Ye-Kitaev model,''
  \href{http://journals.aps.org/prd/abstract/10.1103/PhysRevD.94.106002}{Phys.\ Rev.\ D {\bf 94}, no. 10, 106002 (2016)}
\href{http://arxiv.org/abs/1604.07818}{{\ttfamily arXiv:1604.07818 [hep-th]}}.

\bibitem{Sachdev:1992fk}
S.~Sachdev and J.-w. Ye, ``{Gapless spin fluid ground state in a random,
  quantum Heisenberg magnet},''
  \href{http://dx.doi.org/10.1103/PhysRevLett.70.3339}{Phys. Rev. Lett.
  {\bfseries 70} (1993) 3339},
\href{http://arxiv.org/abs/cond-mat/9212030}{{\ttfamily arXiv:cond-mat/9212030
  [cond-mat]}}.
  
  \bibitem{ParcolletGeorgesnF}
O.~Parcollet and A.~Georges, ``Non-fermi-liquid regime of a doped Mott
  insulator,'' \href{http://dx.doi.org/10.1103/PhysRevB.59.5341}{Phys.
  Rev. B {\bfseries 59} (Feb, 1999) 5341--5360}.
  
  \bibitem{Sachdev:2010um}
S.~Sachdev, ``{Holographic metals and the fractionalized Fermi liquid},''
  \href{http://dx.doi.org/10.1103/PhysRevLett.105.151602}{Phys. Rev.
  Lett. {\bfseries 105} (2010) 151602},
\href{http://arxiv.org/abs/1006.3794}{{\ttfamily arXiv:1006.3794 [hep-th]}}.
 
  \bibitem{ParcolletGeorges}
 O.~Parcollet and A.~Georges, ``Transition from overscreening to underscreening in the multichannel Kondo model: exact solution at large N, " \href{http://journals.aps.org/prl/abstract/10.1103/PhysRevLett.79.4665}{Phys. Rev. Lett.
  {\bfseries 79} (1997) 4665}, \href{http://arxiv.org/abs/cond-mat/9707337}{{\ttfamily arXiv:cond-mat/9707337
  [cond-mat]}}; O.~Parcollet, A.~Georges, G.~Kotliar, and A.~Sengupta, ``{Overscreened
  multichannel SU(N) Kondo model: Large-N solution and conformal field
  theory},''\href{http://journals.aps.org/prb/abstract/10.1103/PhysRevB.58.3794}{Phys. Rev. B {\bfseries 58} (1998) 3794},
\href{http://arxiv.org/abs/cond-mat/9711192}{{\ttfamily arXiv:cond-mat/9711192
  [cond-mat]}}.
  
\bibitem{Jevicki:2016bwu} 
  A.~Jevicki, K.~Suzuki and J.~Yoon,
  ``Bi-Local Holography in the SYK Model,''
  \href{http://link.springer.com/article/10.1007%2FJHEP07%282016%29007}{JHEP {\bf 1607}, 007 (2016)}
  \href{http://arxiv.org/abs/1603.06246}{{\ttfamily arXiv:1603.06246 [hep-th]}}.

\bibitem{Jevicki:2016ito} 
  A.~Jevicki and K.~Suzuki,
  ``Bi-Local Holography in the SYK Model: Perturbations,''
  \href{http://link.springer.com/article/10.1007%2FJHEP11%282016%29046}{JHEP {\bf 1611}, 046 (2016)}
  \href{http://arxiv.org/abs/1608.07567}{{\ttfamily arXiv:1608.07567 [hep-th]}}.

\bibitem{Jensen:2016pah}
  K.~Jensen,
  ``Chaos in AdS$_2$ Holography,''
  \href{http://journals.aps.org/prl/abstract/10.1103/PhysRevLett.117.111601}{Phys.\ Rev.\ Lett.\  {\bf 117}, no. 11, 111601 (2016)}
  \href{http://arxiv.org/abs/1605.06098}{{\ttfamily arXiv:1605.06098 [hep-th]}}.
  
\bibitem{Maldacena:2016upp} 
J.~Maldacena, D.~Stanford and Z.~Yang,
  ``Conformal symmetry and its breaking in two dimensional Nearly Anti-de-Sitter space,''
  \href{http://ptep.oxfordjournals.org/content/2016/12/12C104}{PTEP {\bf 2016}, no. 12, 12C104 (2016)}
  \href{http://arxiv.org/abs/1606.01857}{{\ttfamily arXiv:1606.01857 [hep-th]}}.
  


\bibitem{Engelsoy:2016xyb} 
  J.~Engels\"{o}y, T.~G.~Mertens and H.~Verlinde,
  ``An investigation of AdS$_{2}$ backreaction and holography,''
 \href{http://link.springer.com/article/10.1007%2FJHEP07%282016%29139}{JHEP {\bf 1607}, 139 (2016)}
 \href{http://arxiv.org/abs/1606.03438}{{\ttfamily arXiv:1606.03438 [hep-th]}}.
 
\bibitem{Cvetic:2016eiv} 
  M.~Cvetic and I.~Papadimitriou,
  ``AdS$_{2}$ holographic dictionary,''
 \href{https://link.springer.com/article/10.1007%2FJHEP12%282016%29008}{ JHEP {\bf 1612}, 008 (2016)
  Erratum: [JHEP {\bf 1701}, 120 (2017)]}
 \href{https://arxiv.org/abs/1608.07018}{{\ttfamily [arXiv:1608.07018 [hep-th]]}}.
  
\bibitem{Gu:2016oyy} 
  Y.~Gu, X.~L.~Qi and D.~Stanford,
  ``Local criticality, diffusion and chaos in generalized Sachdev-Ye-Kitaev models,''
  \href{https://arxiv.org/abs/1609.07832}{{\ttfamily arXiv:1609.07832 [hep-th].}}

\bibitem{Gross:2016kjj} 
  D.~J.~Gross and V.~Rosenhaus,
  ``A Generalization of Sachdev-Ye-Kitaev,''
   \href{https://arxiv.org/abs/1610.01569}{{\ttfamily arXiv:1610.01569 [hep-th].}}
  
\bibitem{Berkooz:2016cvq} 
  M.~Berkooz, P.~Narayan, M.~Rozali and J.~Sim\'on,
  ``Higher Dimensional Generalizations of the SYK Model,''
   \href{https://arxiv.org/abs/1610.02422}{{\ttfamily arXiv:1610.02422 [hep-th].}}


\bibitem{Witten:2016iux} 
  E.~Witten,
  ``An SYK-Like Model Without Disorder,''
  \href{https://arxiv.org/abs/1610.09758}{{\ttfamily arXiv:1610.09758 [hep-th].}}

\bibitem{igorgrisha} 
  I.~R.~Klebanov and G.~Tarnopolsky,
  ``Uncolored Random Tensors, Melon Diagrams, and the SYK Models,''
  \href{https://arxiv.org/abs/1611.08915}{{\ttfamily arXiv:1611.08915 [hep-th].}}
  
\bibitem{gs} M.~B.~Green and J.~H.~Schwarz,
  ``Covariant Description of Superstrings,''
  \href{http://www.sciencedirect.com/science/article/pii/0370269384920215}{Phys.\ Lett.\  {\bf 136B}, 367 (1984).}
  


   \bibitem{hansson}
A.~Quelle, T.~Kvorning, T.H.~Hansson, and C.~Morais Smith, 
``Edge Majoranas on locally flat surfaces - the cone and the Mobius band,"
 \href{http://journals.aps.org/prb/abstract/10.1103/PhysRevB.94.125137}{Phys.\ Rev.\ B {\bf 94}, 125137 (2016)}
 \href{https://arxiv.org/abs/1606.06093}{ {\ttfamily arXiv:1606.06093 [cond-mat.str-el].}}
 
  \bibitem{KPZ} 
  V.~G.~Knizhnik, A.~M.~Polyakov and A.~B.~Zamolodchikov,
  ``Fractal Structure of 2D Quantum Gravity,''
 \href{http://www.worldscientific.com/doi/abs/10.1142/S0217732388000982}{ Mod.\ Phys.\ Lett.\ A {\bf 3}, 819 (1988).}
  
   \bibitem{robbert}{R. Dijkgraaf, A geometrical approach to two-dimensional conformal field theory, Ph.D. thesis (Utrecht 1989).}
 
 \bibitem{HV}
  H.~L.~Verlinde,
  ``{Conformal Field Theory, 2-d Quantum Gravity And Quantization Of Teichm\"{u}ller Space,}''
  \href{http://www.sciencedirect.com/science/article/pii/055032139090510K}{Nucl.\ Phys.\ B {\bf 337}, 652 (1990)};  H.~L.~Verlinde and E.~P.~Verlinde,
 ``{Conformal Field Theory And Geometric Quantization,''}
  In *Trieste 1989, Proceedings, Superstrings '89* 422-449 \href{https://lib-extopc.kek.jp/preprints/PDF/2000/0031/0031891.pdf}{{\ttfamily [PUPT-89-1149, IASSNS-HEP-89-58].}} 

\bibitem{WZW-gauged} 
  E.~Witten,
  ``On Holomorphic factorization of WZW and coset models,''
  \href{http://link.springer.com/article/10.1007/BF02099196}{Commun.\ Math.\ Phys.\  {\bf 144}, 189 (1992).}
 
\bibitem{BlauThompson} 
  M.~Blau and G.~Thompson,
  ``Derivation of the Verlinde formula from Chern-Simons theory and the G/G model,''
  \href{http://www.sciencedirect.com/science/article/pii/055032139390538Z}{Nucl.\ Phys.\ B {\bf 408}, 345 (1993)}
 \href{https://arxiv.org/abs/hep-th/9305010v2}{{\ttfamily arXiv:9305010 [hep-th].}}
  
 \bibitem{mps}
 M.~Fannes, B.~Nachtergaele and R.F.~Werner \href{http://link.springer.com/article/10.1007/BF02099178}{Commun. Math. Phys. {\bf 144} (1992) 443}; 
 D. Perez-Garcia, F. Verstraete, M. Wolf, and J. Cirac,
\href{http://www.rintonpress.com/journals/qiconline.html#v7n56}{Quantum Info. Comput. {\bf 7}, 401 (2007)} \href{https://arxiv.org/abs/quant-ph/0608197}{{\ttfamily arXiv:0608197 [quant-ph]}.}


\bibitem{Fu:2016vas} 
  W.~Fu, D.~Gaiotto, J.~Maldacena and S.~Sachdev,
  ``Supersymmetric SYK models,''
  \href{https://arxiv.org/abs/1610.08917}{{\ttfamily arXiv:1610.08917 [hep-th].}}
  
  \bibitem{nyu} 
  S.~Dubovsky, R.~Flauger and V.~Gorbenko,
  ``Solving the Simplest Theory of Quantum Gravity,''
  \href{http://link.springer.com/article/10.1007%2FJHEP09%282012%29133}{JHEP {\bf 1209}, 133 (2012)}
\href{https://arxiv.org/abs/1205.6805}{{\ttfamily arXiv:1205.6805 [hep-th].}}


\bibitem{smirzam} 
  F.~A.~Smirnov and A.~B.~Zamolodchikov,
  ``On space of integrable quantum field theories,''
  \href{https://arxiv.org/abs/1608.05499}{{\ttfamily arXiv:1608.05499 [hep-th].}}  
 
 
  \bibitem{italians} 
  A.~Cavaglià, S.~Negro, I.~M.~Szécsényi and R.~Tateo,
  ``$T \bar{T}$-deformed 2D Quantum Field Theories,''
  \href{http://link.springer.com/article/10.1007%2FJHEP10%282016%29112}{JHEP {\bf 1610}, 112 (2016)}
  \href{https://arxiv.org/abs/1608.05534}{{\ttfamily arXiv:1608.05534 [hep-th].}}
  

\bibitem{ttbar} 
  L.~McGough, M.~Mezei and H.~Verlinde,
  ``Moving the CFT into the bulk with $T\bar T$,''
    \href{https://arxiv.org/abs/1611.03470}{{\ttfamily arXiv:1611.03470 [hep-th].}}
    
    \bibitem{AP}
  A.~Almheiri and J.~Polchinski,
  ``Models of AdS$_{2}$ backreaction and holography,''
  \href{http://link.springer.com/article/10.1007%2FJHEP11%282015%29014}{JHEP {\bf 1511} (2015) 014}
\href{https://arxiv.org/abs/1402.6334}{{\ttfamily arXiv:1402.6334 [hep-th].}}

\bibitem{freidel} 
  L.~Freidel,
  ``Reconstructing AdS/CFT,''
  \href{http://xxx.lanl.gov/abs/0804.0632}{{\ttfamily arXiv:0804.0632 [hep-th].}}
  

\bibitem{Turiaci:2016cvo} 
  G.~Turiaci and H.~Verlinde,
  ``On CFT and Quantum Chaos,''
  \href{http://link.springer.com/article/10.1007%2FJHEP12%282016%29110}{JHEP {\bf 1612}, 110 (2016)}
  \href{http://arxiv.org/abs/1603.03020}{{\ttfamily arXiv:1603.03020 [hep-th].}}
  
  \bibitem{Jackson:2014nla} 
S.~Jackson, L.~McGough, and H.~Verlinde, ``{Conformal Bootstrap, Universality
  and Gravitational Scattering},''
  \href{http://dx.doi.org/10.1016/j.nuclphysb.2015.10.013}{ Nucl. Phys. B {\bfseries 901} (2015) 382--429},
\href{http://arxiv.org/abs/1412.5205}{{\ttfamily arXiv:1412.5205 [hep-th]}}.

   
\end{thebibliography}
\end{document}